\documentclass[10pt,aps,prd,twocolumn,tightenlines,superscriptaddress, letterpaper, amsmath, amssymb, preprintnumbers, floatfix, longbibliography, nofootinbib]{revtex4-2}
\usepackage{graphicx}
\usepackage{amsmath}
\usepackage{physics}
\usepackage[dvipsnames]{xcolor}
\usepackage{wrapfig}
\usepackage{float}
\usepackage{tikz}
\usepackage{array}
\usepackage{qcircuit}
\usepackage{comment}
\usepackage{enumitem}
\usepackage{bm}
\usepackage{tabularx}
\usepackage{multirow}
\usepackage{soul}
\usepackage{orcidlink}
\usepackage{subfig}

\newcommand{\round}[1]{\ensuremath{\left\lfloor#1
\right\rceil}}

\begin{document}

\renewcommand{\thefootnote}{\alph{footnote}}
\title{Practical Application of the Quantum Carleman Lattice Boltzmann Method in Industrial CFD Simulations}

\author{Francesco Turro\,\orcidlink{0000-0002-1107-2873}}
\email{francesco.turro@leonardo.com}
\affiliation{Quantum Computing Research Laboratory,
Leonardo S.p.A., Via R. Pieragostini 80, Genova, Italy}

\author{Alessandra Lignarolo }
\email{alessandra.lignarolo.ext@leonardo.com}
\affiliation{Quantum Computing  Research Laboratory,
Leonardo S.p.A., Via R. Pieragostini 80, Genova, Italy}

\author{Daniele Dragoni\,\orcidlink{0000-0002-1644-5675} }
\email{daniele.dragoni@leonardo.com}
\affiliation{Quantum Computing Research Laboratory, 
Leonardo S.p.A., Via R. Pieragostini 80, Genova, Italy}
\affiliation{Digital Infrastructures, Leonardo S.p.A., Via R. Pieragostini 80, Genova, Italy}

\begin{abstract}
Computational Fluid Dynamics simulations are crucial in industrial applications but require extensive computational resources, particularly for extreme turbulent regimes.  While classical digital approaches remain the standard, quantum computing promises a breakthrough by enabling a more efficient encoding of large-scale simulations with a limited number of qubits.

This work presents a practical numerical assessment of a hybrid quantum-classical approach to CFD based on the Lattice Boltzmann Method (LBM). The inherently non-linear LBM equations are linearized via a Carleman expansion and solved using the quantum Harrow-Hassidim-Lloyd algorithm (HHL). We evaluate this method on three benchmark cases featuring different boundary conditions—periodic, bounce-back, and moving wall—using state-vector emulation on high-performance computing resources.

Our results confirm the validity of the approach, achieving median error fidelities on the order of $10^{-3}$  and success probabilities sufficient for practical quantum state sampling. Notably, the spectral properties of small lattice systems closely approximate those of larger ones, suggesting a pathway to mitigate one of HHL’s bottlenecks: eigenvalue pre-evaluation.

\end{abstract}

\maketitle

\section{Introduction}
Computational Fluid Dynamics (CFD) is a fundamental discipline in fluid mechanics that enables the numerical solution of the Navier-Stokes equations, which govern the motion of fluids. Navier-Stokes are nonlinear partial differential equations of the following form, 
\begin{equation}
    \rho \frac{\partial}{\partial t} u +\rho u\cdot \nabla u = - \nabla P +\nu  \nabla^2 u  + F
\end{equation}
with $u$ being the fluid velocity, $\rho$ the fluid density, $\nu$ the viscosity, $P$ the pressure and $F$ the external force. \cite{McLean2012}
Given their complexity, analytical solutions exist only for highly simplified cases, necessitating numerical techniques for real-world applications, which include the advanced design of planes and cars, the weather forecast prediction, and even the analysis of fluids in biological  tissues \cite{industrial_application_1,industrial_application_2,industrial_application_3,industrial_application_4,industrial_application_5}. 
Modern CFD approaches leverage discretization methods such as finite difference, finite volume, and finite element techniques to transform the continuous equations into solvable algebraic systems. The computational cost of solving these equations grows significantly with the problem size and resolution, making High-Performance Computing (HPC) approached indispensable within industry-relevant settings. 

Despite the computational power of HPC, fundamental limitations remain when tackling extreme-scale problems, such as turbulence at very high Reynolds numbers~\cite{Pope2000,Schlichting2000} 
$Re= \frac{\rho L u}{\nu} \,,$
($L$ being the characteristic length) or multi-physics simulations involving fluid-structure interactions. 

Quantum computing offers a promising avenue to overcome some of these constraints, potentially complementing traditional HPC methods. In fact,  quantum parallelism can be exploited by algorithms that offer a memory scaling with exponential advantage in the number of simulation elements with respect to classical counterparts. Quantum computational fluid dynamics (QCFD) is therefore a promising field of research where CFD simulations are performed according to various schemes directly on quantum devices~\cite{gaitan2020finding,PhysRevResearch.5.033182,meng2024simulating,chen2022quantumfem,oz2022solving,liu2021efficient,gaitan2021finding,li2024potentialquantumadvantagesimulation,Montanaro2016fem,jin2022time}. 

In this work, we tackle prototypical industry-relevant problems within the QCFD setting. Our approach builds upon the well-known Lattice Boltzmann method (LBM) \cite{succi2001lattice, Chen1998} that is a popular CFD technique that describes the temporal evolution of a fluid system on a regular grid, where a statistical distribution represents the probability of the fluid moving in a specific direction. The equations that rule this method are the Lattice Boltzmann equations, from which the Navier-Stokes equations can be obtained under well-established approximations. Specifically, at each time step, this equation alternates between two steps: collision and streaming. The collision step models the effects associated to the internal fluid collisions, while the streaming step propagates the fluid distributions across neighboring lattice points.

Although the Lattice Boltzmann equations are inherently nonlinear, various quantum-based algorithms have been proposed to address their resolution~\cite{itani2023, itani2023quantumalgorithmlatticeboltzmann, Kocherla2024Fullquantumalgorithm, 10.1016/j.jcp.2024.112816, schalkers2024importance, georgescu2024qlbmquantumlattice}.
In this work, to handle the nonlinearity of the collision step, we adopt the Carleman linearization scheme~\cite{Carleman1932}, as discussed in Refs.~\cite{sanavio2024gradapproach, sanavio2025carleman}. 
This method reformulates the original set of nonlinear equations into a linear system by introducing auxiliary variables, at the expense of an approximation error that depends on the truncation order $\nu$.

Following Ref.~\cite{li2024potentialquantumadvantagesimulation}, we rewrite the Carleman Lattice Boltzmann equations in a linear system of the form $A x= b$, where the solution $x$ encodes the fluid distribution for all time steps  and $b$ is connected to the initial fluid distribution. We highlight that the linear set of equations can be solved not only for the evolution of a single time step but also for multiple time steps in one shot, potentially allowing for simulations on longer time scales. Notably, the number of qubits required to encode the $A$ matrix grows logarithmically with the number of time steps, providing favorable scaling for the proposed method. The solution to the linear system is then achieved by implementing the famous HHL algorithm. Although various variants of the HHL have been proposed in the literature
~\cite{zaman2023step,PhysRevLett.110.230501,dervovic2018quantumlinearsystemsalgorithms}, we stick here to the original proposal~\cite{hhl}. The implementation of variational quantum linear solver~\cite{BravoPrieto2023variationalquantum, eni2024} will be explored in future studies.

This work is organized as follows:
Section~\ref{sec:LBM} introduces the LBM and Carleman schemes. Section~\ref{sec:HHL} reviews the implementation of HHL method for solving the Carleman LBM equations. Section~\ref{sec:systems} describes the prototypical industrial use cases investigated in this work, while Sec.~\ref{sec:results} is devoted to the discussion of the numerical results. 
The conclusions are reported in Sec.~\ref{sec:conclusion}.

\section{Lattice Boltzmann method and Carleman linearization scheme \label{sec:LBM}}
\begin{figure}
    \centering
\includegraphics[scale=1.5]{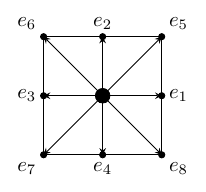}
    \caption{The nine-velocity set of the D2Q9 LBM model. The $e_0$ is the central point of  the lattice.}
    \label{fig:D2Q9}
\end{figure}

\begin{table}[t]
    \centering
    \begin{tabular}{|c|ccccccccc|}
         \hline
         & 0 & 1 &2 &3 &4 &5 & 6&7&8\\
         \hline
         $e_i$ & (0,0) & (1,0) &(0,1) &(-1,0) & (0,-1) & (1,1) & (-1,1)&(-1,-1)&(1,-1)\\
         $w_i$& $\frac{4}{9}$ & $\frac{1}{9}$ &$\frac{1}{9}$ &$\frac{1}{9}$ &$\frac{1}{9}$ &$\frac{1}{36}$ & $\frac{1}{36}$&$\frac{1}{36}$&$\frac{1}{36}$\\
         \hline
    \end{tabular}
    \caption{Values for the D2Q9 Lattice Boltzmann Method used in this work.}
    \label{tab:D2Q9}

\end{table}

The LBM~\cite{succi2001lattice} relies on the discretization of space and time on a regular lattice with spacing $dx$ and $dt$, respectively. For simplicity, in this work, we adopt $dx=dt=1$. On such lattices, fluid dynamics is described using discrete-velocity distribution functions $f_i(t,n)$, which represent the statistical densities of fictitious fluid particles moving along a specific direction  $e_i$ at lattice site $n=(n_x,n_y)$ and time $t$. Specifically, we employ the two-dimensional D2Q9 discretization, where the fluid velocity vectors $e_i$ are defined as shown in Fig.~\ref{fig:D2Q9}.
From these distribution functions, key macroscopic fluid properties such as density $\rho$ and the fluid velocity $\vec{u}=(u_x,\,u_y)$ can be recovered at each lattice point
$n$ as follows:
\begin{equation}
    \rho(n,t) = \sum_{i=0}^{8} f_i(t,n) ,
\end{equation}
\begin{equation}
    u_{j}(n,t) = \frac{1}{\rho} \sum_{i=0}^8 e_i\cdot e_{j} f_i(t,n)\,.  \label{eq:u_velocity}
\end{equation}

The evolution of the distribution functions is governed by the Lattice Boltzmann Equation (LBE):
\begin{equation}\label{eq:LBE}
f_i(t+dt,n+e_i)-f_i(t,n) = -\Omega(f_i(t,n)) .  
\end{equation}
Here, the left-hand side describes the streaming of particles at site $n$ along the direction $e_i$, while the right-hand side models the collisional interactions of the fluid particles at the same site. A widely used approach for the collision operator is the Bhatnagar-Gross-Krook (BGK) approximation~\cite{Bhatnagar1954}, which assumes
\begin{equation}
    \Omega_i(f_i(t,n)) = - \omega (f_i(t,n) - f_i^{eq}(t,n)) \,,
\end{equation}
where $\omega=\frac{dt}{\tau}$ is the relaxation parameter, related to the fluid relaxation time$\tau$, epresents the Maxwell-Boltzmann equilibrium distribution function. This equilibrium distribution is given by
\begin{equation}\label{eq:equil}
    f^{eq}_i(t,n)= w_i \rho(t,n) \left(1+
    \frac{\vec{u} \cdot e_i}{c_s^2}+\frac{(\vec{u}\cdot e_i)^2}{2 c_s^4}- \frac{\vec{u}\cdot \vec{u}}{2 c_s^2}\right)\,,
\end{equation}
where $c_s=\frac{1}{\sqrt{3}} \frac{dx}{dt}=\frac{1}{\sqrt{3}}$ is the lattice sound speed, $\vec{u}=u(t,n)$ is is the fluid velocity from Eq.\eqref{eq:u_velocity}, and the weights $\omega_i$ and the velocity set $e_i$ for D2Q9 are provided in Tab.~\ref{tab:D2Q9}.

Under the BGK approximation, the LBE simplifies to
\begin{equation}
    f_i(t+dt,n+e_i)= - \omega (f_i(t,n) - f_i^{eq}(t,n))+f_i(t,n)\,.
\end{equation}
To explicitly separate the collision and streaming steps, we introduce the auxiliary function $f_i^*(t,n) $, allowing the LBE to be rewritten as
\begin{align} 
\text{Collision:} \quad & f_i^*(t,n) = (1-\omega) f_i(t,n) + \omega f^{eq}_i(t,n) , \\
\text{Streaming:} \quad & f_i(t+dt,n+e_i) = f_i^*(t,n) . \end{align} 
This formulation highlights that the streaming process remains linear in $f_i(t,n)$, whereas the collision term introduces nonlinearity due to Eq.~\ref{eq:equil}.\\

As demonstrated in Ref.~\cite{Sanavio2024,itani2023}, the nonlinear LBM equations can be reformulated into a system of linear equations, making them more amenable to quantum computing approaches. This transformation is achieved through the application of the Carleman linearization scheme~\cite{Carleman1932,Sanavio2024,sanavio2024gradapproach,sanavio2025carleman}, which introduces auxiliary variables defined as
\begin{align}
g_{ij}(t,n,m)&=f_i(t,n)\,f_j(t,m), \\
    h_{ijk}(t,n,m,l)&=f_i(t,n)\,f_j(t,m)\,f_k(t,l) ,
\end{align}
representing all possible products of one, two, ..., up to  $\nu$ distribution functions $f_i$, where $\nu$ denotes the Carleman truncation order. A detailed formulation of the Carleman streaming operator is provided in App.~\ref{app:Carleman_formula}. 
By incorporating higher-order terms, this approach allows for a more accurate representation of the nonlinear dynamics.

Fundamentally, the Carleman linearization scheme can be interpreted as a Taylor series expansion of the LBM equations, truncated at order $\nu$. For instance, under this formulation, the collision term in the LBM equation transforms into:
\begin{equation}
\begin{split}
f_i^*(t,n)= &\sum_{j}^Q \frac{\partial f_i^*(t,n) } {\partial f_j(t,n)} f_j(t,n) + \\
&\sum_{j,k}^Q \frac{\partial^2 f_i^*(t,n) } {\partial f_j(t,n) \partial  f_k(t,n)} f_j(t,n) f_k(t,n) + ...  \\
     = &\sum_{j}^Q E_{ij} f_j(t,n) + \sum_{j,k}^Q E_{ijk} g_{jk}(t,n,m) + ... \; . \\
\end{split}   
\end{equation}    

However, the auxiliary variables $g(t,n,m), h(t,n,m,l)$,  and higher-order terms must also be updated according to their definitions. Specifically, for a second-order truncation ($\nu=2$), where $f$ and $g$ variables are retained, the transformed equations take the form:
\begin{align}\label{eq:collision_carleman}
    f_i^*(t,n)  = \sum_j D_{ij} \,f_j(t,n) + \sum_{jk} E_{ijk}\, g_{jk} (t,n,m), \\ 
    g_{rs}^*(t,n,m)  = \sum_{i,j,k,l} D_{ri} D_{sj}\,  g_{ij}(t,n,m)\,,    
\label{eq:gcollision_carleman}
\end{align}
where $n\,,m \in [0,...,N]$ with $N=N_x\,N_y$. The explicit expressions for the matrices $D$ and $E$ are provided in App.~\ref{app:Carleman_formula}.

To express Eq.~\ref{eq:gcollision_carleman} in a more compact form, we rewrite it in matrix notation as
\begin{equation}
  \begin{pmatrix}
        F^*(t)\\
        G^*(t)\\
    \end{pmatrix}
=     \begin{pmatrix}
       D & E \\
        0 & D\otimes D\\
    \end{pmatrix}
    \begin{pmatrix}
   F(t)\\
G(t)
\end{pmatrix},
\label{eq:carleman_collision}
\end{equation}
where the state vector $\phi(t)=(F(t), G(t))^T$ collects all auxiliary variables. Specifically, the vectors are defined as $F(t)^T = (f_{0}(t,0), ..., f_8(t,N-1))$, and $G(t)^T=(g_{00}(t,0),...,g_{88}(t, (N-1) (N-1)))^T$. This matrix formulation enables a structured representation of the Carleman linearization, facilitating its implementation in quantum algorithms.

A similar expansion can be applied to the streaming step, yielding:
\begin{equation}
  \begin{pmatrix}
        F(t+dt)\\
        G(t+dt)\\
    \end{pmatrix}
=     \begin{pmatrix}
       S & 0 \\
        0 & S \otimes S \\
    \end{pmatrix}
    \begin{pmatrix}
   F^*(t)\\
G^*(t) \label{eq:straming_carleman}
\end{pmatrix}\,, 
\end{equation}
where $S$ represents the streaming matrix (see App.~\ref{app:Carleman_formula} for details). 
Since the streaming equation remains linear in the
$f^*$ components, even under standard boundary conditions, the Carleman expansion preserves this linearity. For a generic truncation order $\nu$, the updated matrix is given by the tensor product of $\nu$ copies of $S$, denoted as $\otimes_\nu S$. 

Consequently, at an arbitrary time $t$, the LBM equations in their Carleman-linearized form can be expressed as
\begin{equation}
\phi(t+dt) = C \phi(t)\,, \label{eq:LBM_Carleman}
 \end{equation}
where $\phi(t)$ is the state vector encompassing al $\nu$ Carleman variables, and $C$, known as the Carleman matrix, is obtained by multiplying the matrices from Eqs.~\eqref{eq:carleman_collision}, \eqref{eq:straming_carleman}:
\begin{equation}
   C= \begin{pmatrix}
        S  & 0 \\
        0 & S \otimes S \\
    \end{pmatrix} \begin{pmatrix}
       D &E  \\
        0 & D \otimes D \\
    \end{pmatrix}\,. \label{eq:Carleman_full}
\end{equation}\\

In classical computation, the applicability of this method is significantly constrained by the rapid growth of the system’s matrix size, which scales as $(NQ)^{\nu}$, where $N$ is the number of lattice points, and $Q$ the number of fluid components per lattice site. In contrast, in a quantum computing framework, the number of qubits $n_s$ required to encode these matrices scales more favorably as $n_s \sim \nu \log(QN)$, suggesting a potential computational advantage. However, a major challenge arises from the fact that the Carleman matrix $C$ is not unitary. cannot be directly implemented on a quantum computer, preventing its direct implementation on a quantum computer. The following section explains how to address this limitation.

\section{From Carleman to the Quantum HHL algorithm\label{sec:HHL}}

To solve the Carleman equation over $N_t$ time steps, starting from an initial time $t_0$ with the initial  condition $\phi(t_0)=\phi_0$,  we express the set of $N_t$ Carleman equations as:
\begin{align}
   \phi(t_0) &= \phi_0 \nonumber\\
    \phi(t_0+dt) - C \phi(t_0)& =0 \nonumber\\
    ...&\nonumber\\
    \phi(t_0+N_t dt) - C \phi(t_0+(N_t-1)dt)& =0  \, .     
\end{align}
Here, we iteratively apply Eq.~\eqref{eq:LBM_Carleman} over multiple time steps.

To efficiently encode all solutions of these Carleman equations, we introduce the vector $x$ as:
\begin{equation}
    x=\left( \phi(t_0), \phi(t_0+dt), ...,\phi(t_0+N_t\,dt) \right) ^T \,.
\end{equation}
Following Ref.~\cite{li2024potentialquantumadvantagesimulation}, we can write the Carleman equations in a compact linear system of form
\begin{equation}
\Tilde{A} x = b\,,
\end{equation}
where  the matrix $\Tilde{A}$ is defined as:
\begin{equation}
\Tilde{A}=    \begin{pmatrix}
        1 & 0 & 0 &...&0\\
        -C & 1 & 0 &...&0\\
        0 & -C & 1 &...&0\\
        ... &  ... & ... &...&0\\
        ... & 0 & 0 &...&1\\
    \end{pmatrix}\,, \label{eq:carlemanA}
    \end{equation}
and the right-hand side vector $b$ is given by:
\begin{equation}
b= \begin{pmatrix}
    \phi_0, & 0, & ..., & 0\\
\end{pmatrix}^T\,. \label{eq:Carleman_linear_system}
\end{equation}
Each entry of the solution vector $x$ corresponds to the evolution of the system at a given time step $t_i$. For example, from the first row, we retrieve the initial condition:
\begin{equation} x(0) = \phi_0. \end{equation} From the second row, we obtain:
\begin{equation} x(t_0+dt) = C x(t_0) = C \phi_0 = \phi(t_0+dt), \end{equation} which directly follows from Eq.~\eqref{eq:LBM_Carleman}. This structure allows us to iteratively reconstruct the system's evolution while leveraging the Carleman linearization framework.

Solving this system of linear equations can be efficiently achieved using the HHL algorithm~\cite{hhl}.
Given a Hermitian matrix $A$ and a vector $\vec{b}$, the HHL algorithm computes a solution vector $\vec{x}$ such that $A\vec{x} = \vec{b}$.  Although the $\tilde{A}$ matrix in Eq.~\eqref{eq:carlemanA} is not inherently hermitian, it is possible to construct an equivalent Hermitian representation as:
\begin{equation}
 A   = \begin{pmatrix}
     0 & \Tilde{A}^\dagger \\
      \Tilde{A} & 0 \\
 \end{pmatrix}  \label{eq:A_HHL}.
\end{equation}
To ensure consistency with the Carleman-LBE formulation, the linear system then takes the form:
\begin{equation}
    A \begin{pmatrix}
        x(t) \\
        0\\
        \end{pmatrix}
   =\begin{pmatrix}
        0 \\
        b\\
    \end{pmatrix} =\vec{b}. 
\end{equation}

Notably, the computational cost of evolving the system over $N_t$ time sttime steps scales logarithmically with the number of qubits, i.e., $n_s \sim \log_2(N_t + 1)$.
This logarithmic scaling offers significant advantages in both storage efficiency and the simulation of large lattices over extended time periods, as the required memory grows only logarithmically with system size. Consequently, this approach enables the simulation of later time steps with a minimal quantum resource overhead.

\subsection{HHL algorithm}

\begin{figure}[t]
    \centering
    \includegraphics[width=1\columnwidth]{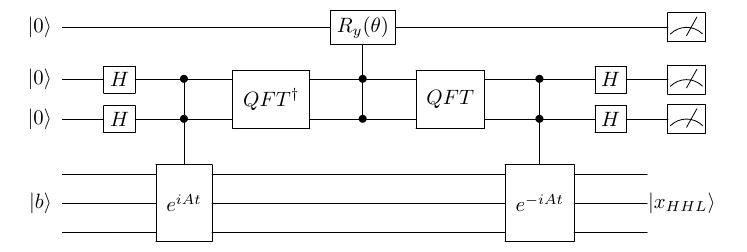}
    \caption{Scheme of HHL circuit. The top qubit corresponds to the ancilla qubit, the second and third qubits represents the clock qubits, and the bottom register (forth, fifth and sixth qubits) encodes the linear system in the system qubit. A detailed description of the single gates is reported in the main text.}
    \label{fig:HHL_circuit}
\end{figure}

This section provides a concise review of the key components of the Harrow-Hassidim-Lloyd (HHL) algorithm~\cite{hhl,zaman2023step,PhysRevLett.110.230501,dervovic2018quantumlinearsystemsalgorithms,tsemo2024enhancingharrowhassidimlloydhhlalgorithm}. 
The HHL algorithm employs two main registers: a clock register with $n_c$ qubits responsible for the numerical precision of the solution $x$, and a system register, with $n_b$ qubits, encoding the details of the actual system problem through $A$ and $b$. Additionally, an extra ancilla qubit is used during the computation.
The algorithm consists of five fundamental steps, illustrated in Fig.\ref{fig:HHL_circuit}:
\begin{enumerate}
    \item \textbf{State preparation.} Encode the vector $\vec{b}$ into the system register while initializing all other qubits in state $\ket{0}$.
    \item \textbf{Quantum Phase Estimation (QPE)~\cite{Berry2006,Nielsen_Chuang_2010}.} Apply Hardamard gates to the clock qubits. Implement controlled unitary operations $e^{i t_i A}$, for specific times $t_i$, with control over the clock register (Hamiltonian simulation). In the end, implement the adjoint Quantum Fourier Transform (QFT) on the clock qubits.     
    \item \textbf{Controlled rotations.} Apply  $R_y$ rotations to the ancilla qubit with the rotation angles $\theta$ determined by the closest binary approximation $\tilde{\lambda}$ of the eigenvalue $\lambda$ of the matrix $A$.
    
    \item \textbf{Inverse QPE.} Revert the Quantum Phase Estimation.\\
    \item \textbf{Measurement.} Measure the ancilla and clock registers. If the ancilla qubit is in $\ket{1}$ state and the clock register in $\ket{0...0}$ state, the system register contains a binary approximation of $\vec{x}$ with the desired accuracy.   
\end{enumerate}

The successful execution of the HHL algorithm hinges on two crucial factors: the efficient encoding of the initial state and the accurate implementation of the controlled multi-qubit evolution $e^{iAt}$. In this study, we emulate the quantum circuit classically using the Davinci1 HPC system\footnote{We use 48 AMD EPYC Rome 7402 CPU processors with NVIDIA A100 GPUs of the Leonardo S.p.A. company's Davinci-1 computers.}, where both state initialization and Hamiltonian simulation are computed exactly, without relying on gate decomposition.

In App.~\ref{app:numbergats}, we provide an upper bound on the number of CNOT gates required for compiling the Hamiltonian simulation, showing that it scales linearly with the number of lattice points and quadratically with the number of time steps.

A fundamental requirement for the correctness of HHL is that the number of clock qubits $n_c$ must be $n_c \geq n_{c}^{min}$, where   $n_c^{min}=\log_2\left(\lceil \frac{\lambda_{max}}{\lambda_{min}} \rceil \right)$. Here $\lceil \cdot \rceil$ denotes the ceiling function,  $\lambda_{min} (\lambda_{max})$ indicates the smallest (largest) positive eigenvalue of $A$. This condition ensures that all binary representations of eigenvalues of $A$ are different from zero.

The time evolution parameter $t_i$ in the Hamiltonian simulation is chosen as
\begin{equation}
    t=2^{{n_i}} \,2 \pi \frac{1}{4\lambda_{max}}\,.
\end{equation}
where $n_i$ indexes the clock qubits ($n_i=0,...,n_c-1$). 
This choice allows the QPE circuit to separate binary representations of positive and negative eigenvalues. Specifically, for any eigenvalue $\lambda$,  the clock qubits would be $\frac{\lambda}{4}$ when $\lambda>0$ and $1+\frac{\lambda}{4}$ otherwise. The factor $1/4$ resolves degeneracies associated with $\pm \lambda_{max}$, a refinement necessary for robust eigenvalue separation. 

This study employes an approximate version of the HHL method by implementing a limited number of controlled $R_y$-rotations. The corresponding angles are determined from 
\begin{equation}
    \theta=\arcsin\left(\frac{C_p}{\bar{\lambda}}\right) \label{eq:Cry_angles}
\end{equation}
where $C_p$ is a tunable parameter, set to 1 in this work, and $\bar{\lambda}$ represents the binary approximation of  $\lambda$, given by
\begin{equation}
    \bar{\lambda} = \round{{2^{n_c}}\frac{\lambda}{4 \lambda_{max}}}\,.
\end{equation}
Here, $\lambda_{max}$ denotes the maximum eigenvalue, and $\round{\cdot}$ rounds to the nearest integer. 

In this work, we compute the actual eigenvalue spectrum, but as discussed in Sec.~\ref{sec:HHL_spectra},  it is possible to estimate the spectrum of a larger lattice systems using those of a smaller ones. This technique is explained in Sec.~\ref{sec:different_spectra}, where we examine its applications to fluid dynamics simulations. If the spectrum is unknown, a complete implementation of HHL requires covering the entire eigenvalue range, necessitating $2^{n_c}$ different controlled rotations. This exponential growth in required operations presents a significant challenge for near-term quantum hardware, reinforcing the importance of spectrum estimation techniques and hybrid classical-quantum approaches. 

In the literature, one finds that the original time complexity of HHL is given by $O(\kappa^2 s^2 \log(n)/\epsilon)$~\cite{hhl} where $n$ indicates the dimension of the matrix (in our case, $n\sim Q\,L$),  $\kappa$ the condition number, and $s$ is the sparsity of the linear matrix. For our studied system, Carleman at the first order, the sparsity decreases linearly with $L$ because we have $L$ small block-matrices, therefore, $s=\frac{L}{L^2} s_b$ where $s_b$ indicates the sparsity of the smallest block matrix. Different works present improvements of the HHL \cite{HHL_improvement1,PhysRevLett.120.050502}. The best scaling also for dense matrix is given by $O(\kappa^2 polylog(n) ||A||/\epsilon)$~\cite{PhysRevLett.120.050502} where $||\,\cdot\,||$ indicates the Frobenius norm.

\begin{figure*}[t!]
    \centering
\includegraphics[width=1.0\textwidth]{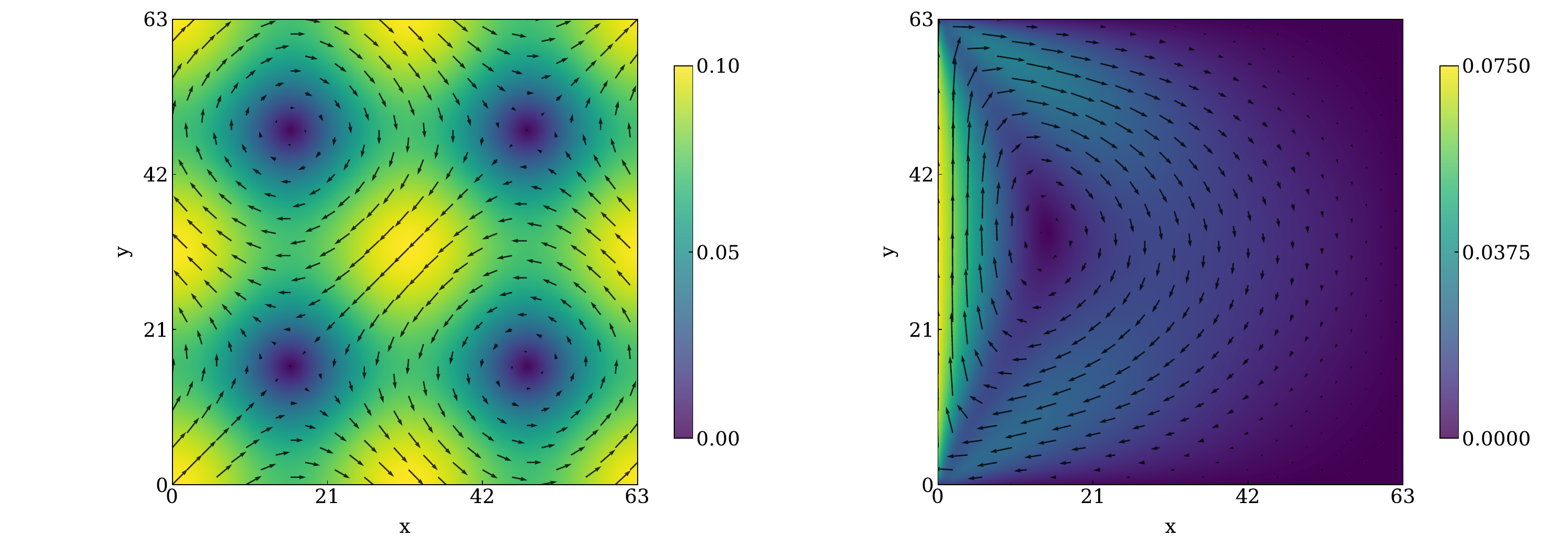}
    \caption{ (Left Panel) Steady State solution of the Kolmogorov flow using parameters reported in Tab.~\ref{tab:parameters}. (Right Panel) Lid-driven asymptotic solution for a square lattice of size $N_x=64$ with $v_{lid}=(0,0.075)$ at $Re=25$.}
    \label{fig:lid_driven_result}
\end{figure*}

\section{Use Cases\label{sec:systems}}

We validate the proposed method by applying it to three simplified use cases that serve as proxies for prototypical industrial applications. 

The first system represents an open flow configuration, where the periodic boundary conditions (PBC) are defined to simulate an unbounded domain. The second system models a confined flow, incorporating a no-slip boundary condition implemented via a bounce-back scheme, which ensures elastic scattering of fluid particles at the wall.
In both cases, the initial fluid distribution follows a Kolmogorov flow profile, characterized by a sinusoidal distribution field defined as:
\begin{equation}
f_i(n_x,n_y) = \omega_i \left[ 1+ A_x \cos(\frac{2\pi}{N_y} k_x y)+ A_y \cos(\frac{2\pi}{N_x} k_y x)\right] \label{eq:kolmogorov}
\end{equation}
where we set the parameters $A_x=A_y$ and $k_x=k_y=1$. This setup induces a well-defined initial velocity distribution, providing a suitable testbed for evaluating the performance of our method under different flow conditions. The left panel of Fig.~\ref{fig:lid_driven_result} illustrates the initial velocity field of the Kolmogorov flow for a square lattice of size $N_x=64$.
The third system under consideration is the lid-driven cavity flow, an extension of the bounce-back case in which one or more walls move at a constant velocity $\vec{v}_{lid}$. In this work, we spcecifically analyze a configuration where the left wall moves with a velocity of $(0,v_{lid})$. The initial state consists of a quiescent fluid with zero velocity throughout the domain, except for the nodes adjacent to the moving wall, where the velocity is set to match. The right panel of Fig.~\ref{fig:lid_driven_result} presents the velocity field for the evolved lid-driven system of a square lattice of size $N_x=64$, with $Re=25$ and $v_{lid}=0.075$. The simulation results confirm the formation of a characteristic primary vortex induced by the moving lid.

Further details on the boundary conditions and the initial fluid distribution are provided in App.~\ref{app:BC}.
Unless otherwise specified in the text, the initial fluid parameters used in the simulations are listed in Tab.~\ref{tab:parameters}. 

\begin{table}[th]
    \centering
    \begin{tabular}{|c|c|c|}
    \hline
    & PBC/bounce-back & Lid-driven  \\
    \hline
    $A_x$  & 0.3 & 0.0 \\
    $A_y$  & 0.3 & 0.225 \\
    $k_x$  & 1 & 0\\
    $k_y$  & 1 & 0 \\
    $v_{lid}$ & 0.0 & $0.075$\\
    \hline
    \end{tabular}
    \caption{Parameters used to initialize the simulation of the various systems via Eq.~\eqref{eq:kolmogorov}. For the lid-driven cavity case, Eq.~\eqref{eq:kolmogorov} is used only for points next to the moving wall.}
    \label{tab:parameters}
\end{table}

\section{Results \label{sec:results}}

In this section, we apply the proposed methodology to the use cases introduced in previous sections and analyze the corresponding results. The discussion is structured as follows:
In Sec.~\ref{sec:carleman_erorr} we investigate the systematic errors introduced by the Carleman linearization scheme with respect to LBM solutions.
In Sec.~\ref{sec:HHL_spectra} we examine the spectral properties of the $A$  matrix with respect to the lattice sizes. In Sec.~\ref{sec:singletimestep} and Sec.~\ref{sec:multitime} we evaluate the performance of the proposed methods, when evolving the system for a single step and over multiple time steps, respectively. Finally, in Sec.~\ref{sec:different_spectra}, we investigate the accuracy and limitations of using precomputed spectral information for different lattice sizes, assessing its impact on the proposed method's performance.

\subsection{Carleman Error \label{sec:carleman_erorr}}

We evaluate the error introduced by the Carleman linearization scheme in comparison with the LBM implementation. To quantify this error, we compute the Root Mean Square Error (RMSE)~\cite{Sanavio2024} defined as
\begin{equation}
    RMSE(t)= \frac{1}{Q} \sum_{i=0}^{Q}\sqrt{\frac{1}{N_x N_y} \sum_{n}^{N_x N_y} \left(1-\frac{f_{Carleman}(t,n,i)}{f_{LBM}(t,n,i)}\right)^2} \,,
\end{equation}
where $f_{Carleman}(t,n,i)$ and $f_{LBM}(t,n,i)$ indicate the fluid distribution obtained from Carleman and LBM evolution, respectively, at time $t$, for lattice point $n$ and velocity component $i$.

\begin{figure}[t]
    \centering
\includegraphics[width=1.\columnwidth]{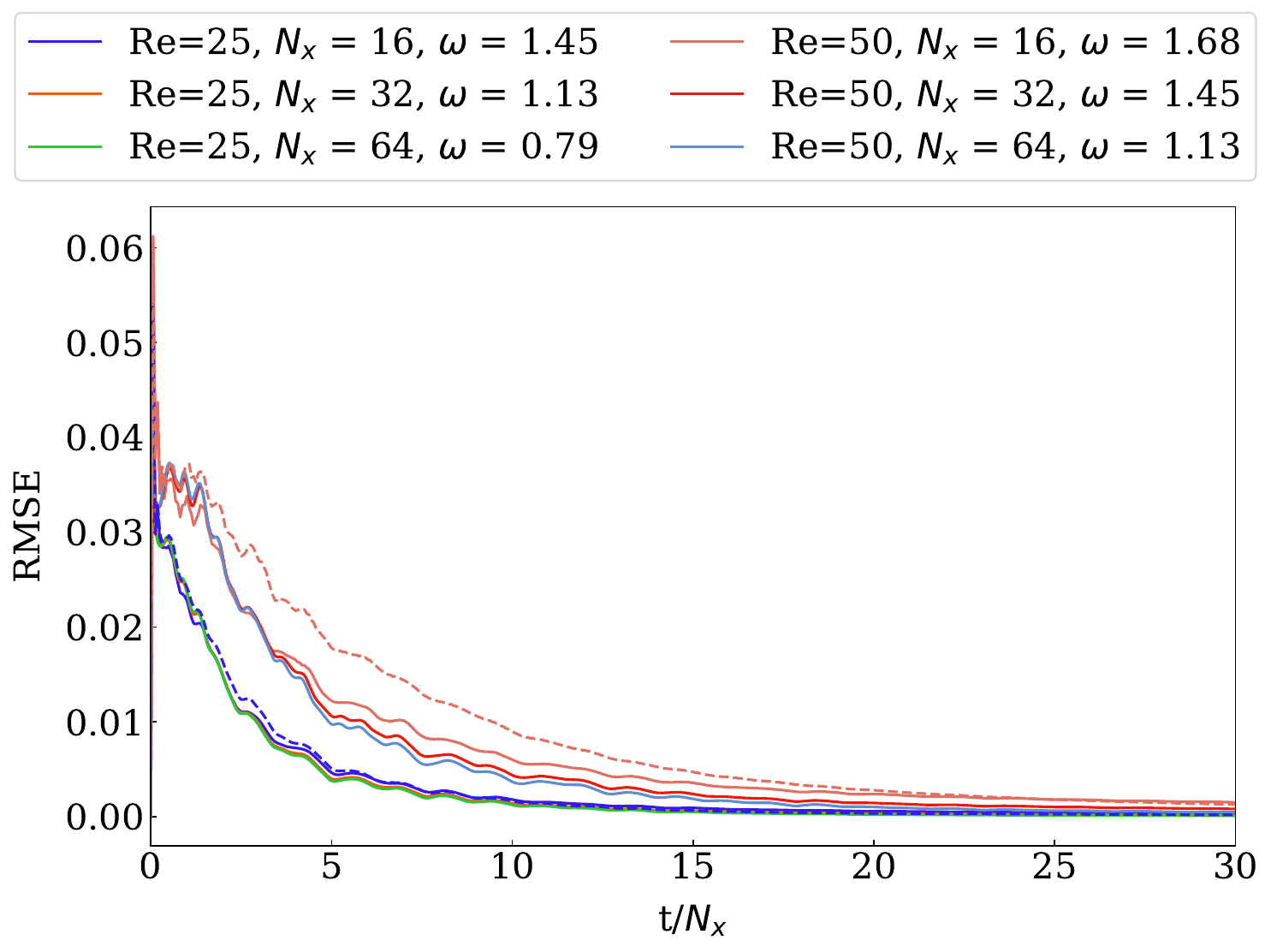}
    \caption{RMSE for the bounce-back conditions as a function of time at various $Re$ and $N_x$ values. Solid lines correspond to first-order Carleman approximation, while dashed line indicate second-order approximation.}
    \label{fig:carleman_RMSE_bounceback}
\end{figure}

Fig.~\ref{fig:carleman_RMSE_bounceback} shows the RMSE results for the bounce-back boundary condition fixing the Reynolds number for different lattice sizes and using the Kolmogorov flow as initial condition. We differentiate between first-order and second-order approximations of the Carleman method: solid lines correspond to first-order simulations, while dashed lines represent second-order results. 

Two key observations emerge from the results. First, all RMSE curves exhibit a decreasing trend towards zero over time. Second, the differences between first- and second-order simulations remain small.  Notably, the second-order Carleman method yields slightly better accuracy for small values of $\omega$, whereas for larger $\omega$,  the first-order approach performs comparably or even slightly better. At $\omega\sim 1$ instead, first- and second-order RMSE are very similar.
On the contrary, in the case of PBC simulations, the second-order method underperforms relative to the first-order method in the high-$\omega$ regime.

\begin{figure}[t]
    \centering
    \includegraphics[width=1.\columnwidth]{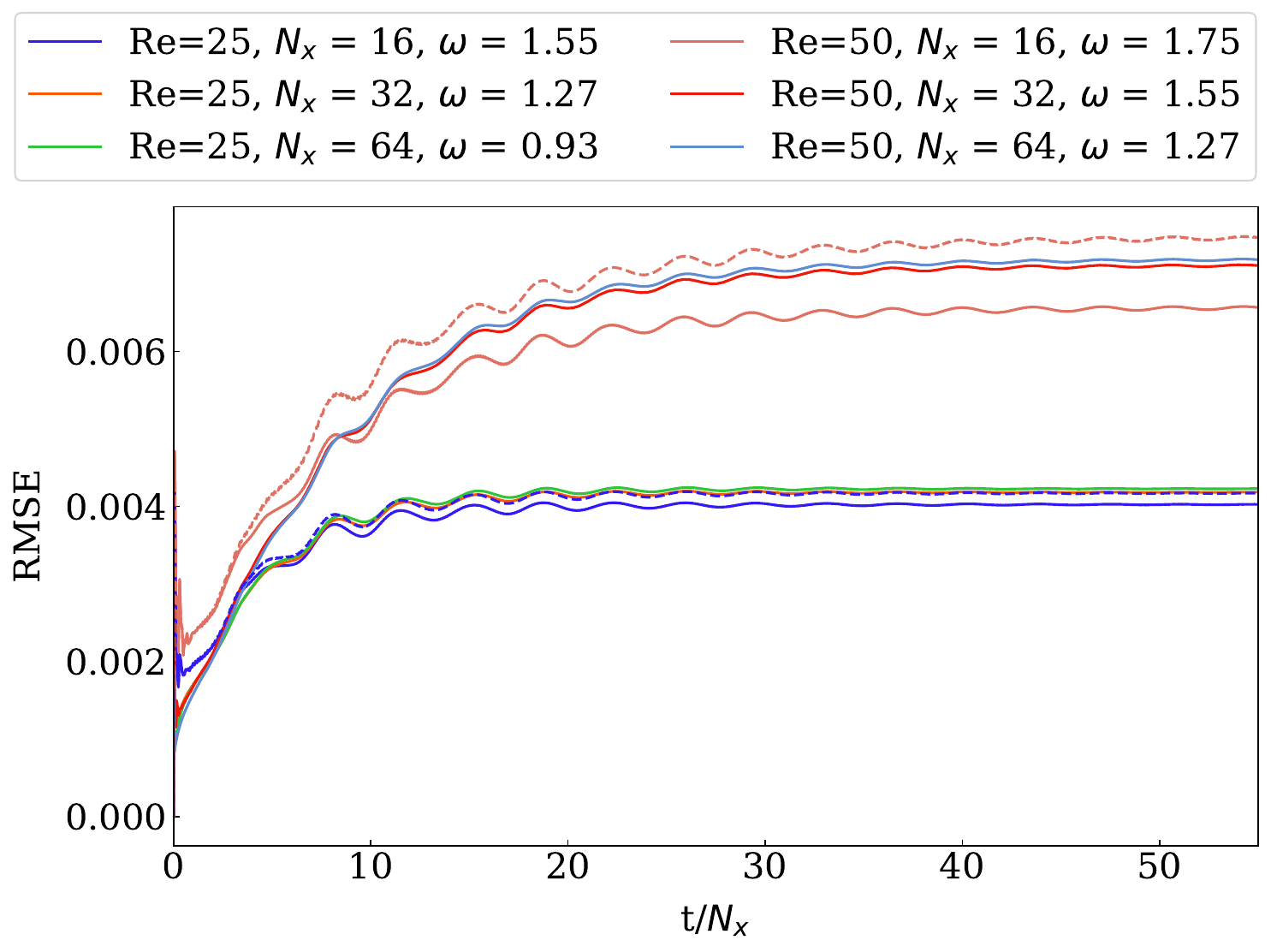}
     \caption{RMSE for the lid-driven system as a function of time at various $Re$ and $N_x$ values. Solid lines correspond to first-order Carleman approximation, while dashed line indicate second-order approximation.}
    \label{fig:carleman_RMSE_lid}
\end{figure}

For the lid-driven cavity problem (see App.\ref{app:lid_driven} for details).  the RMSE results are displayed in Fig.~\ref{fig:carleman_RMSE_lid}. We use the same legend as in Fig.~\ref{fig:carleman_RMSE_bounceback}. Unlike in previous cases, the RMSE does not decay to zero but instead stabilizes at a plateau, with an average error of approximately $0.6\%$ at late times. This persistent error predominantly arises in regions where vortex structures form (see Fig.~\ref{fig:lid_driven_result}), highlighting the Carleman linearization’s inherent limitations in capturing turbulent behavior. Surprisingly, for high $\omega$ values, the second-order method performs worse than the first-order approximation.

Based on these findings, we conclude that the first-order Carleman method provides comparable accuracy to the second-order approach while significantly reducing computational overhead. Consequently, we adopt the first-order approximation in our subsequent implementations. This choice enables us to efficiently simulate larger lattice sizes, which would otherwise be infeasible when combining HHL and second-order Carleman linearization due to computational memory constraints.

\subsection{Spectra of $A$ \label{sec:HHL_spectra}}

We identify an interesting characteristic of the Carleman-linearized LBM equation solved using the HHL method. When fixing the number of time steps, the eigenspectra corresponding to different lattice sizes exhibit strong similarities.

\begin{figure*}[t]
\centering
\captionsetup{width=0.8\columnwidth}
\subfloat[Spectra of $A$ for the bounce-back B.C.  for the evolution of a single time step. \label{fig:Bounceback_spectra_T_1}]{\includegraphics[width=0.95\columnwidth]{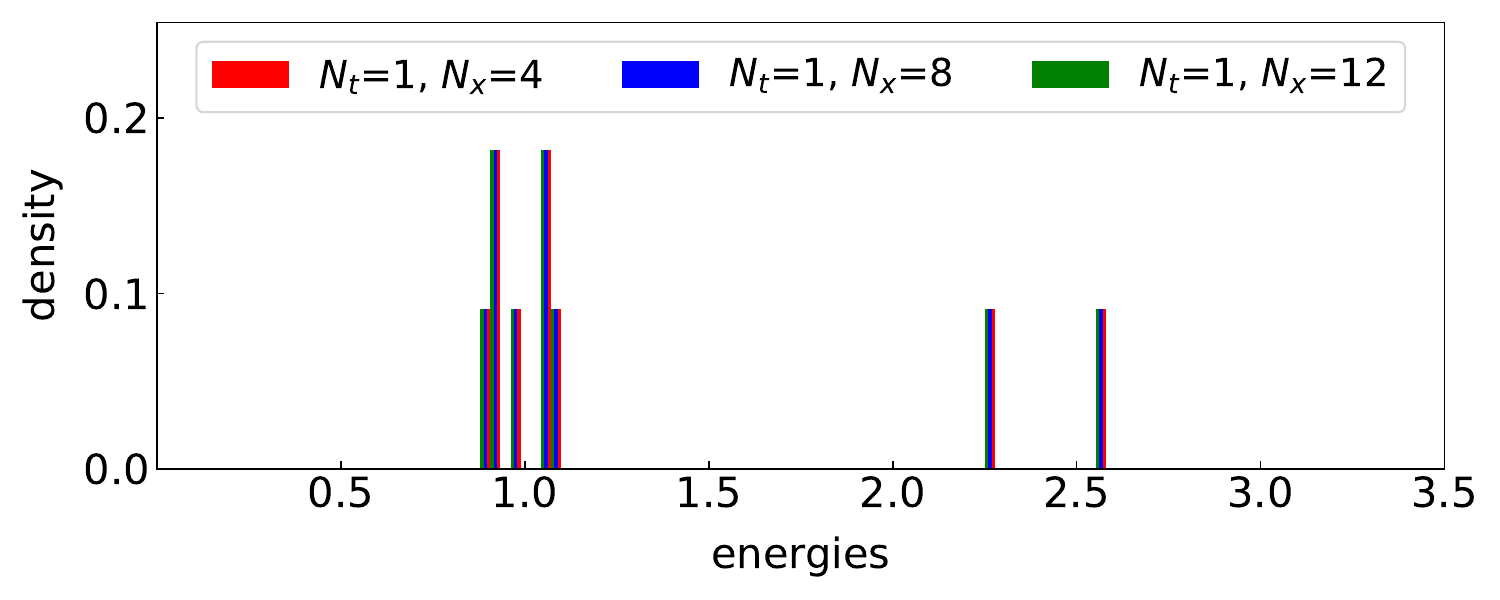}}
\subfloat[Spectra of $A$ for the lid-driven system  for the evolution of a single time step. \label{fig:Lid_spectra_T_1}]
{\includegraphics[width=0.95\columnwidth]{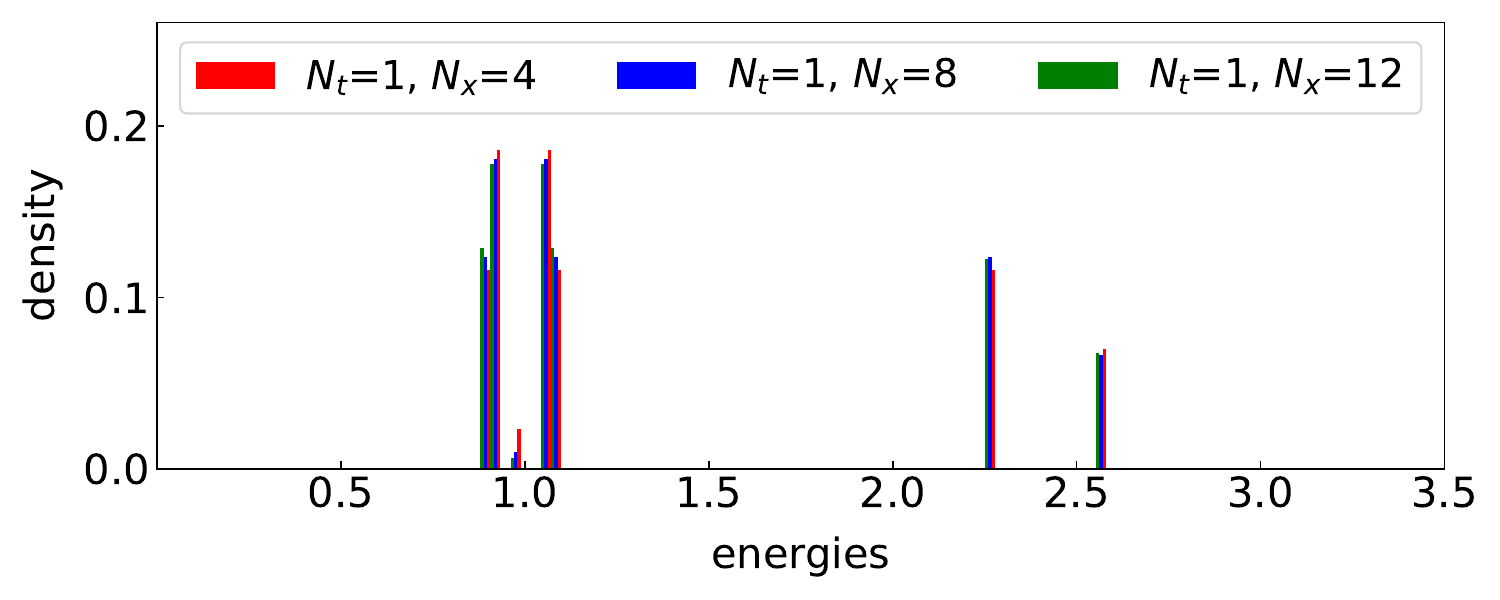}}\\

\subfloat[Spectra of $A$ for the bounce-back B.C.  after evolving for 7 time steps.]
{\includegraphics[width=0.95\columnwidth]{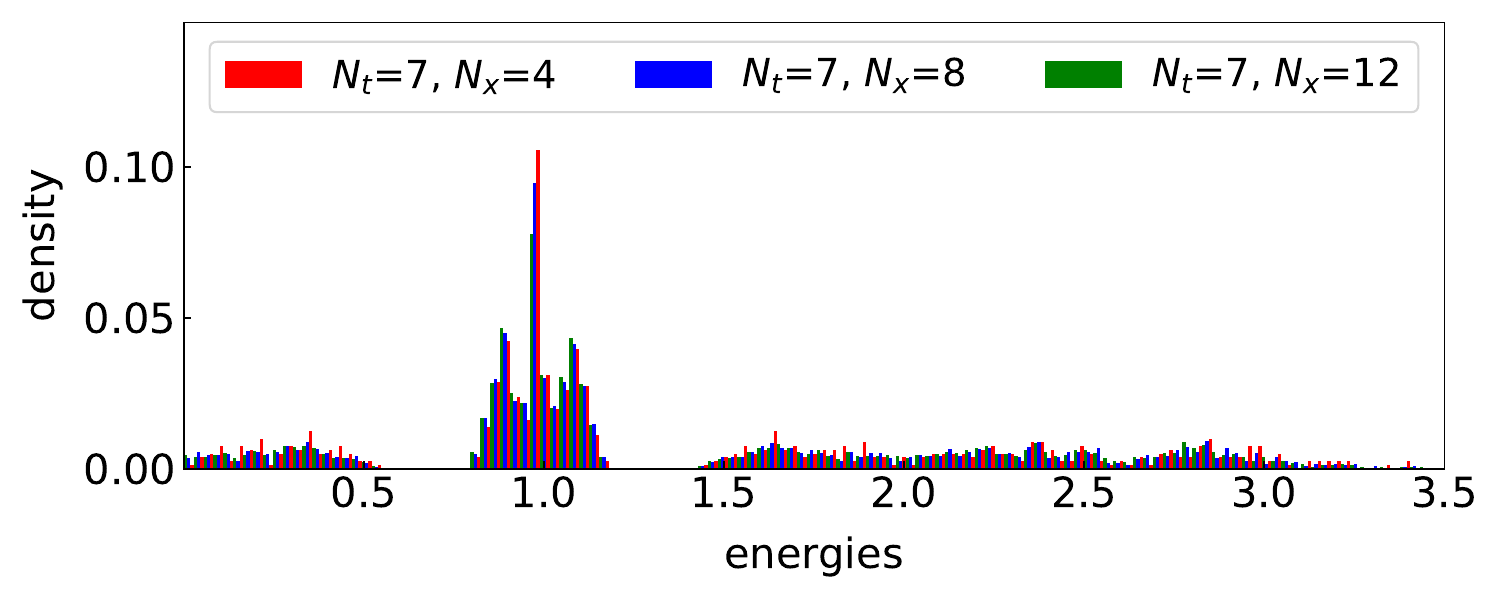}
\label{fig:Bounceback_spectra_T_3}}
\subfloat[Spectra of $A$ for the lid driven system  for the evolution of 7 time steps. 
\label{fig:Lid_spectra_T_3}]{\includegraphics[width=0.95\columnwidth]{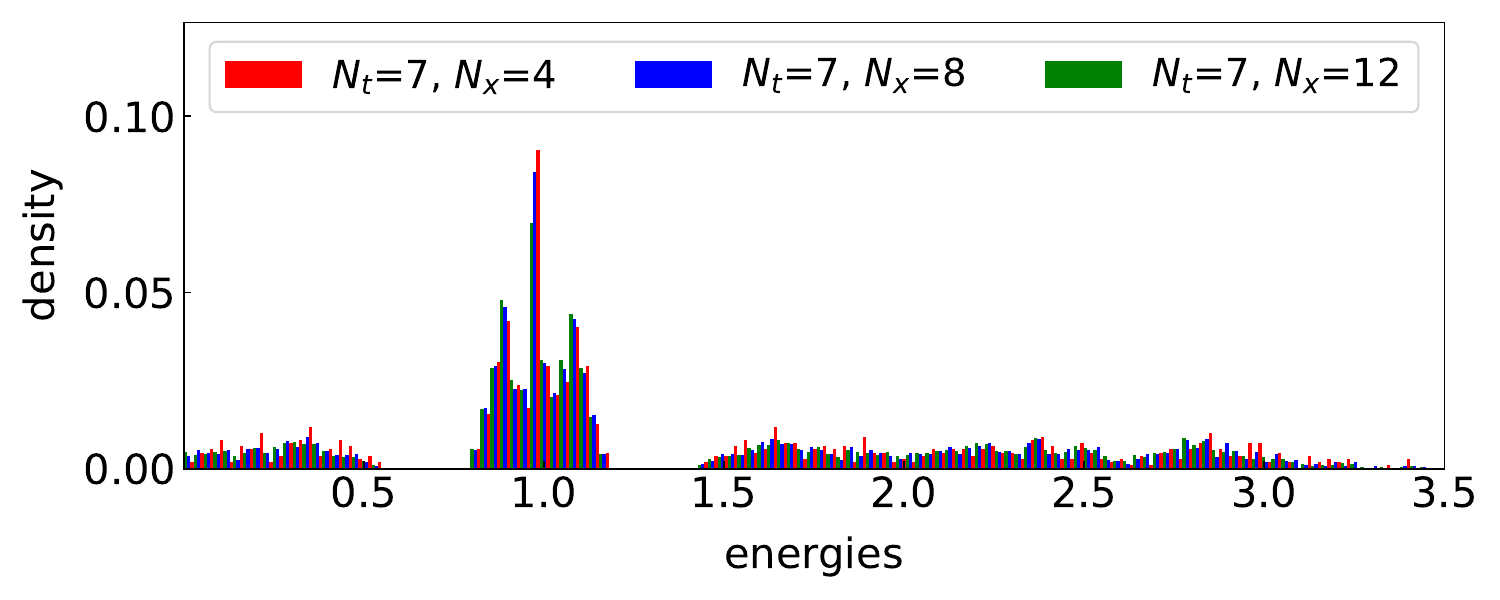}}
\captionsetup{width=\textwidth}
\caption{Positive part of the eigenspectrum of $A$ matrices for different square lattice sizes  $N_x=4,8,16$ and two evolution times $\Delta T$, with $\omega=1.10$. Left panels (a,c) correspond to the bounce-back system, while right panels (b,d) correspond to the lid-driven system. The top panels display single time-step evolution, whereas the bottom panels show multi time-step evolution. In each plot, the bin width is set to $3.5/2^7\sim0.027$.}
\end{figure*}

Figures~\ref{fig:Bounceback_spectra_T_1} and \ref{fig:Lid_spectra_T_1} display the positive half of the eigenspectrum of the matrix $A$\footnote{the full spectrum is symmetric about zero.} for the bounce-back and lid-driven cavity cases (with $v_{lid}=0.075$), respectively, after a single time step, for various lattice sizes. Notably, the spectra remain identical for $N_x=4,8,12$, suggesting that the time evolution of larger lattice systems can be effectively studied using the spectra of smaller ones, which are computationally more tractable to diagonalize.

\begin{figure}
    \centering
    \includegraphics[width=1\columnwidth]{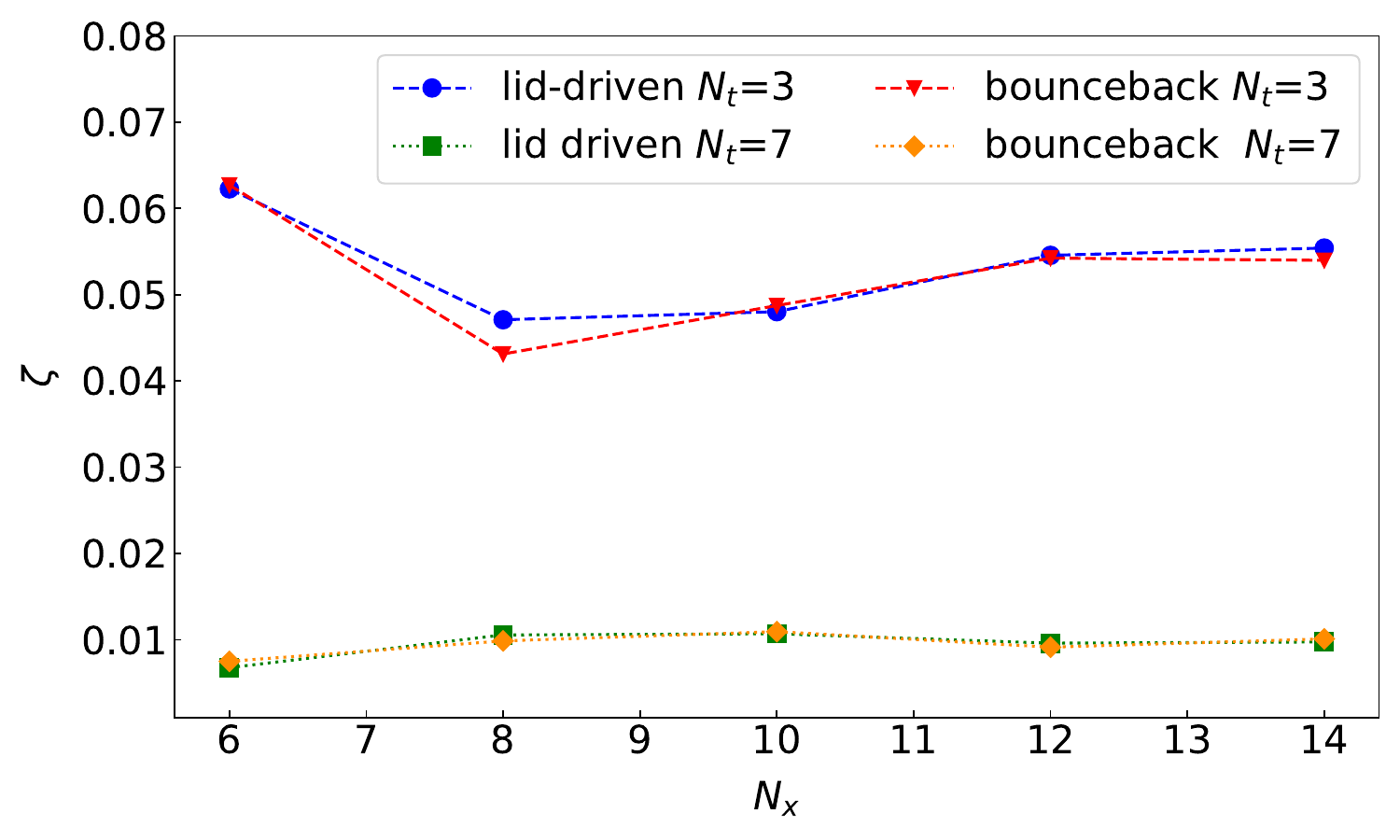}
    \caption{Deviation from the $N_4$ eigenspectrum as a function of $N_x$ calculated via \( \zeta \) function, for \( N_T = 3, 7 \) and two use cases: bounce-back and lid-driven.}
    \label{fig:spectrum_Nx}
\end{figure}

A similar trend is observed when evolving the system over multiple time steps. Figures~\ref{fig:Bounceback_spectra_T_3} and \ref{fig:Lid_spectra_T_3} illustrate the spectral densities after seven time steps. While the spectra for different lattice sizes remain comparable, they are no longer identical. To quantify the deviation between the spectra of larger lattices and that of a smaller

nce lattice \( N_x = 4 \), we define \( \zeta \) as the ratio between the number of counts in the spectrum of $(N_x)$-lattice whose bins are not present in the spectrum of the lattice with \( N_x = 4 \), and the total number of counts:
\begin{equation}
    \zeta = \frac{\sum N_{cs} [\lambda(N_x) \neq \lambda(4)]}{ \sum N_{cs}(\lambda(N_x)]} \,, \label{eq:zeta_counts}
\end{equation}

where $N_{cs}$ represents the number of counts in the bin $\lambda(N_x)$ for a lattice of size $N_x$.
Figure~\ref{fig:spectrum_Nx} shows the computed \( \zeta \) values for different use cases as a function of \( N_x \), considering evolutions over \( N_t = 3 \) and \( N_t = 7 \) time steps. The results indicate that \( \zeta \) remains approximately constant for different lattice sizes. Specifically, for $N_t=1$, all points staisfy $\zeta = 0$; for $N_t=3$, the spectrum discrepancy is around $5\%$, and for $N_t = 7$, it decreases to approximately $1\%$. Moreover, both the lid-driven and bounce-back cases exhibit similar trends.

In Sec.~\ref{sec:different_spectra}, we leverage this property to approximate the spectra of a lattice of size $N_x$ using that of a smaller lattice, reducing computational complexity while maintaining accuracy.

\subsection{Single Time Step Evolution\label{sec:singletimestep}}
We study the performance of the proposed algorithm, which combines the first-order LBM Carleman linearization scheme with the HHL quantum algorithm, by evolving the systems introduced in Sec.~\ref{sec:carleman_erorr} for a single time step.  To compare the performance and reliability of the proposed method, we compute the error fidelity between the analytical evolution computed classically (by applying the Carleman method to the initial state) and the quantum circuit results. The error fidelity is defined as
\begin{equation}
   \epsilon = 1-\left| \braket{\psi}{x_{HHL}}\right|^2\,,
\end{equation}
where $\ket{\psi}$ represents the state that encodes the exact solution obtained from the Carleman scheme and $\ket{x_{HHL}}$ is the state obtained from the HHL algorithm, conditioned on measuring the ancilla qubit in $\ket{1}$ and the clock quibts in $\ket{0...0}$. 

Moreover, we evaluate the success probability to determine the practical feasibility of the method, ensuring that the correct state can be sampled with a fair number of shots. We compute two success probabilities: the first measures the probability of the ancilla qubit being in $\ket{1}$, while the second assesses the probability of obtaining the correct state, where both the ancilla is in $\ket{1}$ and the clock qubits are in $\ket{0000}$.
As the number of clock qubits increases, the first probability is expected to converge to the second, further validating the consistency of the approach.

\begin{figure*}[t]
\centering
\includegraphics[width=0.8\textwidth]{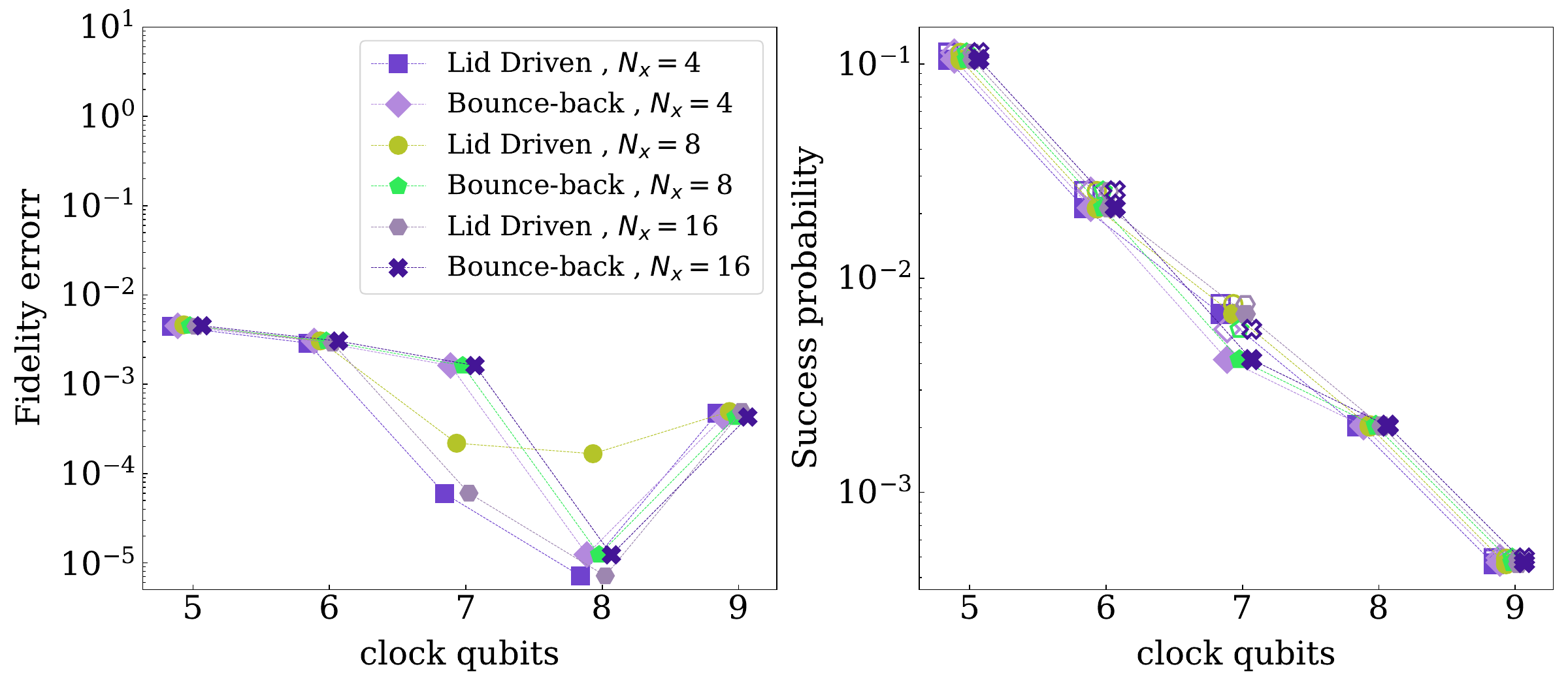}
\caption{Fidelity error (left panel) and success probability (right panel) as functions of the number of clock qubits for different studied systems.  All computations are performed with $\omega=1.1$. In both panels, filled symbols represent the fidelity error and the success probability of measuring the ancilla in $\ket{1}$ with the
clock qubits in the $\ket{0...0}$ state.  Empty symbols in the right panel correspond to the probability of measuring only the ancilla in  $\ket{1}$. The bounce-back system is evolved from $t_0=0$, while the lid-driven cavity system starts from $t_0=40$ (see main text for details).}
\label{fig:clock_qubits_single}
\end{figure*}

We analyze the dependence of the HHL method's performance on the number of clock qubits. Figure~\ref{fig:clock_qubits_single} shows the results for different systems (PBC, bounce-back, and lid-driven cavity) across various lattice sizes. In the case of the lid-driven cavity system, we initiate the evolution from  $t_0 = 40$\footnote{The initial fluid distribution is obtained by applying the Carleman matrix $t_0$ times to the lid-driven cavity initial conditions}, since at time $t_0 \sim 0$, the evolution is closely resembles that of the bounce-back system. Notably, at $t_0 = 40$, a non-linear vortex structure has emerged. As anticipated from HHL theory, increasing the number of clock qubits enhances the accuracy of the solution, bringing it closer to the exact solution of the linear system. However, this improvement comes at the cost of a steep decline in success probability, which decreases exponentially with the number of qubits.  To balance accuracy and practical feasibility for industrial applications, we identify an optimal trade-off: using 6–7 clock qubits ensures high accuracy (error $<10^{-2}$) while maintaining a reasonable success probability ($>10^{-3}$). Notably, for $\geq 9$ clock qubits, numerical errors cause the fidelity error to increase, highlighting one of the difficulties associated with the HHL algorithm.

We further investigate the HHL method's performance as a function of lattice size $N_x$, keeping all other parameters fixed. Our results indicate that both success probability and fidelity remain constant across different lattice sizes. This finding suggests that, as system size increases, the primary computational cost arises from the additional system qubits and quantum gates. Details can be found in App.~\ref{app:size_study}.

Additionally, we assess the robustness of proposed method at fixed Reynolds numbers $Re = 10, 50$, while varying $N_x$ for the three considered systems: PBC, bounce-back, and lid-driven cavity. As before, for the lid-driven cavity evolution is initiated at $t_0=40$.

\begin{figure*}[t]
\centering
\includegraphics[width=0.8\textwidth]{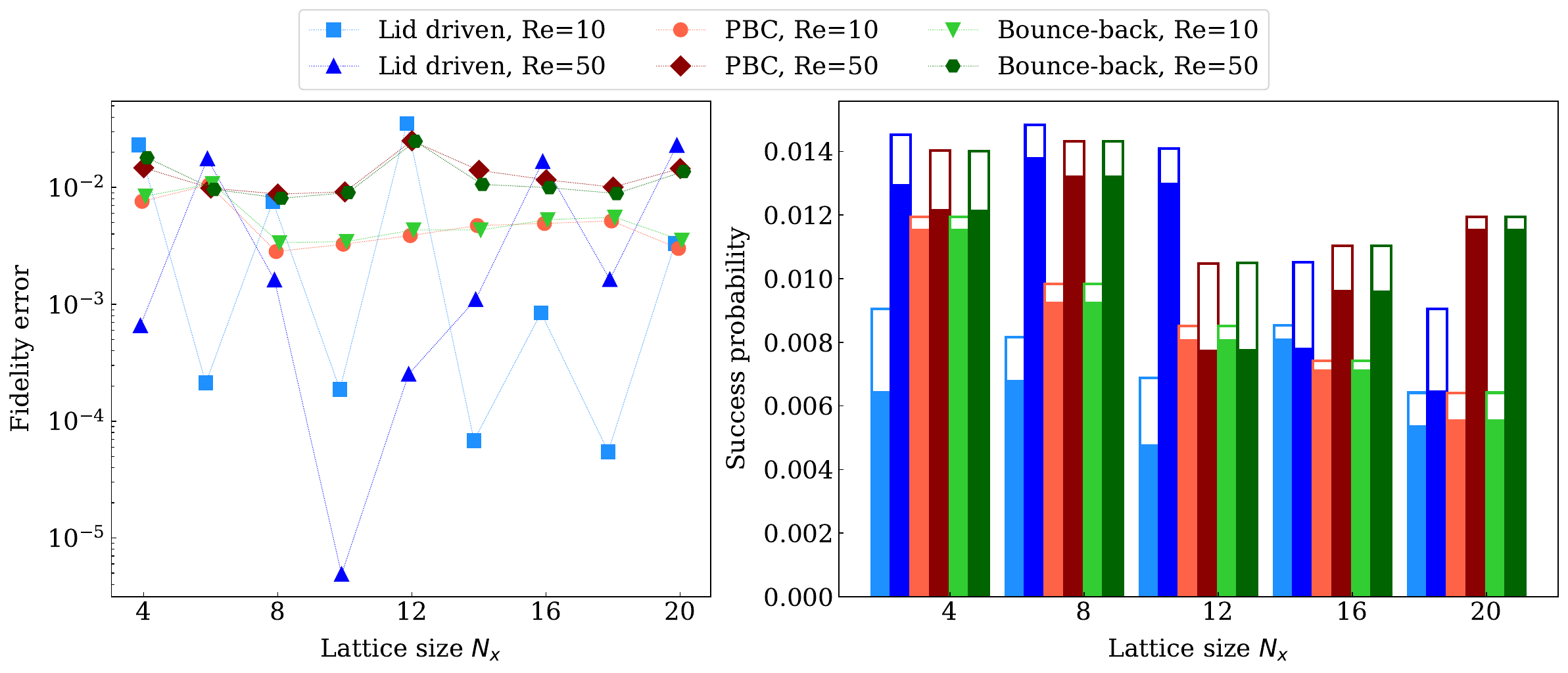}
\caption{Quantum fidelity error (left panel) and success probabilities (right panel) as functions of the lattice size $N_x$ for different systems and Reynolds numbers $Re$. In the left panel, data points are slightly shifted to improve readability and distinguish individual points more clearly. In the right panel, filled symbols represent the success probability of measuring the ancilla in $\ket{1}$ with the clock qubits in the $\ket{0...0}$ state, 
while empty symbols correspond to measuring only the ancilla in $\ket{1}$. The bounce-back boundary conditions are evaluated at $t_0=0$, while the lid-driven cavity system is analyzed starting from $t_0=40$. We set $n_c=7$.}
\label{fig:Reynolds_step1}
\end{figure*}

The left panel of Fig.~\ref{fig:Reynolds_step1} presents the quantum fidelity error, while the right panel illustrates the success probability. Different markers correspond to distinct Reynolds numbers and systems. In the success probability plot, empty bars represent cases where only the ancilla ris measured in $\ket{1}$ state, whereas filled bars correspond to instances where both the ancilla is in $\ket{1}$ and the clock qubits are in $\ket{0...0}$. 

Our findings reveal that, for PBC and bounce-back systems, fidelity error remains largely independent of lattice size, while success probability exhibits variation. I In contrast, for the lid-driven cavity system, we observe no clear trend, as illustrated in Fig.~\ref{fig:size_step1}. This suggests that the combined HHL-Carleman approach is highly sensitive to the choice of $\omega$ and system-specific dynamics. Nonetheless, the maximum observed error in the computed solution remains on the order of $10^{-2}$,  confirming the method’s viability for practical applications.

\subsection{Multi-time steps \label{sec:multitime}}
We extend the proposed method to study system evolution over multiple time steps by incorporating additional rows in Eq.~\ref{eq:A_HHL}. As in the single-step case, we evaluate the quantum fidelity relative to the exact solution at each time step, along with the success probability. The method is implemented for a square lattice of size $N_x = 8$,  initializing the evolution at different starting times $t_0$ to capture distinct stages of fluid evolution: the initial fluid configuration ($t_0 = 0$), an intermediate state ($t_0 = 20$), and the state approaching the asymptotic solution ($t_0 = 40$).

\begin{figure*}[t]
\centering
\subfloat[Obtained results for a $8\times8$ square with P.B.C. for $\omega=1.1,\,1.5$\label{fig:evolution_8x8_pbc}]{\includegraphics[width=\textwidth]{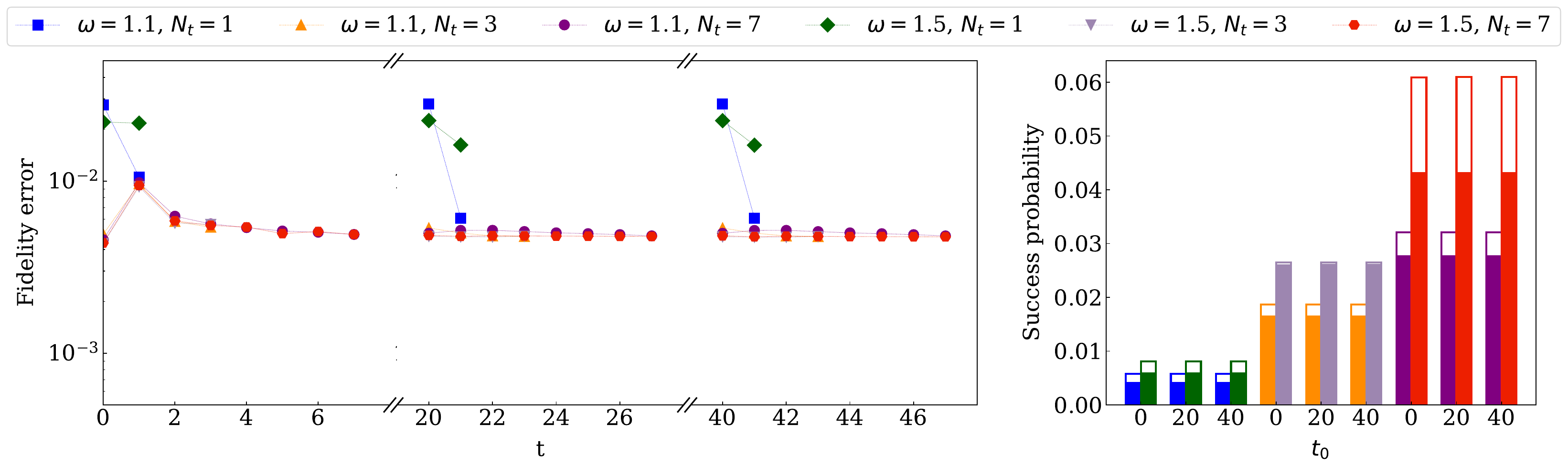}}\\
\subfloat[Obtained results for a $8\times8$ closed box with bounce-back B.C. for $\omega=1.1,\,1.5$.\label{fig:evolution_8x8_bounceback}]{\includegraphics[width=\textwidth]{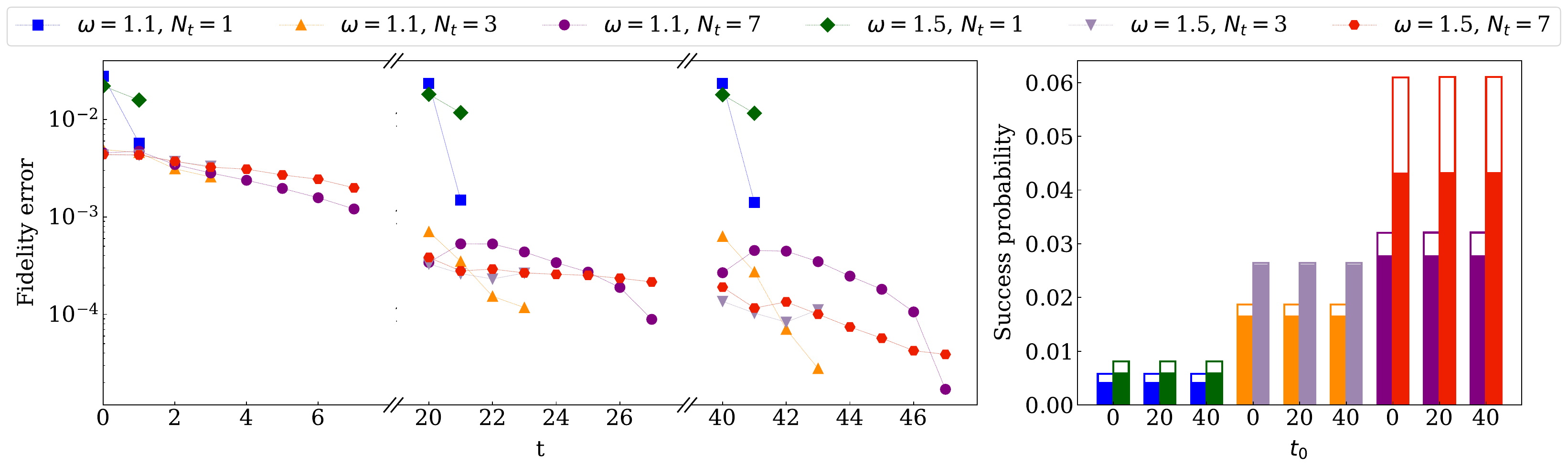}}\\
\subfloat[Obtained results for lid-driven cavity system for a $8\times8$ lattice with $\omega=1.1,\,1.5$. \label{fig:evolution_8x8_lid}]{\includegraphics[width=\textwidth]{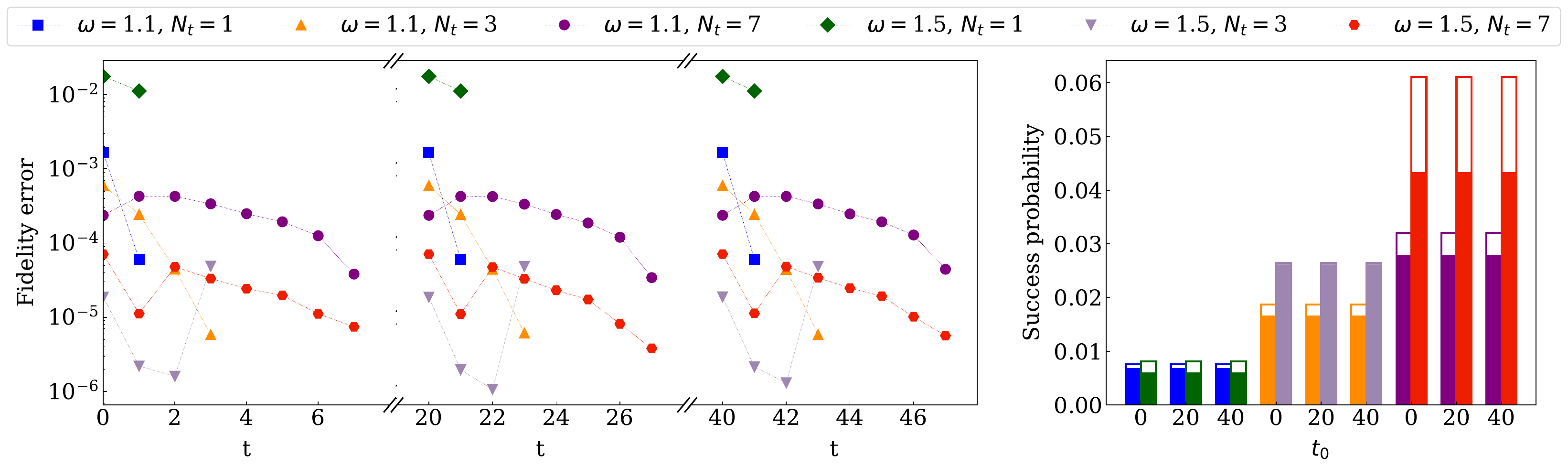}}

\caption{Results of the quantum evolution for $1,3,7$  time steps across different systems (PBC, bounce-back, and lid-driven cavity) on an $8\times8$ lattice, with relaxation frequencies $\omega=1.1,\,1.5$.  Different symbols represent different time steps and values of $\omega$, with $n_c=7$ clock qubits. The left panels display the quantum fidelity error, while the right panels show the corresponding success probabilities. Filled bars represent the probability of measuring the ancilla in $\ket{1}$ and the clock register in the $\ket{0...0}$ state, whereas empty bars indicate only the probability of measuring the ancilla in $\ket{1}$.\label{fig:evolution_8x8}}

\end{figure*}

Figure~\ref{fig:evolution_8x8} 
presents the results for the three studied systems—PBC, bounce-back, and lid-driven cavity (from top to bottom)—with parameter choices of
$\omega = 1.1, 1.5$ and $n_c = 7$.

The results are shown for different evolution times 
$N_t=1,3,7$, corresponding to the use of 1, 2, and 3 qubits for time encoding, respectively. The left panel displays the quantum fidelity error, while the right panel shows the total success probability. For clarity, we use the same colored symbols or bars to represent the results obtained for different initial times ($t_0=0,20,40$), though these calculations are independent, as example, the results at 
$t_=20$ do not depend on those at $t_0=0$.
We also plot the HHL solution for the initial state (shown at $t=t_0=0,20,40$), which often exhibits the highest errors. We can discard the subset of solutions for $t=t_0$, since the initial fluid distribution is already known and serves as a parameter in the HHL method.
We observe that the fidelity error remains relatively stable ($10^{-3}-10^{-2})$ across different systems and different initial times $t_0$. For the bounceback and PBC, the fidelity error decreases across different initial times, also because the fluid is dissipating (it reaches the zero-velocity configuration). For the lid-driven cavity, the fidelity error remains within the same order ($10^{-4}-10^{-3}$) for different initial times, as the fluid converges to a vortex configuration.
Instead, the success probability increases with the number of time steps, suggesting an improvement in the likelihood of obtaining a valid quantum solution as more time steps are included. Interestingly, for shorter evolution times, the deviation from the exact solution is more pronounced.

When the number of qubits is fixed for the HHL method at a specific number of time steps, $N_t$, one can iterate the same quantum circuit $i$-times to propagate for longer times, $i\,N_t$. However, the success probability decreases with each iteration. Based on our results, we suggest that increasing the number of qubits to encode longer times is favorable, as the success probability either remains at the same order or increases.

\begin{figure*}[t]
\centering
\includegraphics[width=1.\textwidth]{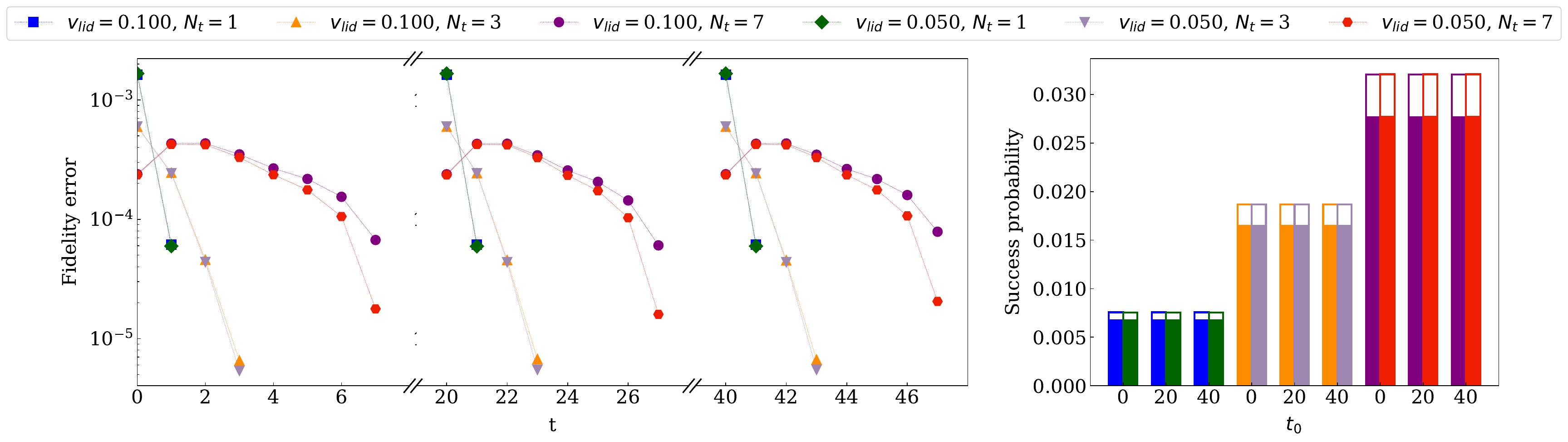}
\caption{Results for different Mach numbers in the lid-driven cavity system across various time evolutions, with $n_c=7$ and $\omega=1.1$. Different symbols correspond to different time steps  ($N_T=1,3,7$) and initial lid velocities. The left panels display the quantum fidelity error, while the right panels show the corresponding success probabilities. Filled bars represent the probability of measuring the ancilla in $\ket{1}$ and the clock register in the 
$\ket{0...0}$ state,  whereas empty bars indicate only the probability of measuring the ancilla in $\ket{1}$.}
\label{fig:vlid_driven}
\end{figure*}

We further analyze the dependence of the proposed algorithm on the Mach number in the lid-driven cavity system. To this end, we consider different initial lid velocities, $v_{lid}=0.05,\,0.10,$, while maintaining the same computational setup as in Fig.~\ref{fig:evolution_8x8_lid}. Figure~\ref{fig:vlid_driven} presents the results for different time steps, with the left and right panels displaying the fidelity error and success probability, respectively, following the same structure as in Fig.~\ref{fig:evolution_8x8}. The results indicate that the Mach number does not significantly impact the performance of the method, as the observed fidelity errors and success probabilities remain nearly identical across different values of $v_{lid}$.

Additionally, we vary the coefficient  $C_p$, which scales the rotation angles in the controlled-$R_y$ gates. As expected, we find that the success probability exhibits a linear dependence on this coefficient. Further details can be found in App.~\ref{app:success_vs_cp}.

\subsection{Simulations via smaller lattice spectra\label{sec:different_spectra}}

In Sec.~\ref{sec:HHL_spectra}, we showed that the eigenvalues of a lattice with a generic size 
$N_x$ closely resemble those of a smaller lattice with $N_x=4$. 
Building on this observation, this section explores the application of the hybrid method when approximating the spectrum of a larger $N_x$ lattice using that of a smaller $N_x=4$ lattice.

\begin{figure*}[t]
    \centering
\includegraphics[width=1\linewidth]{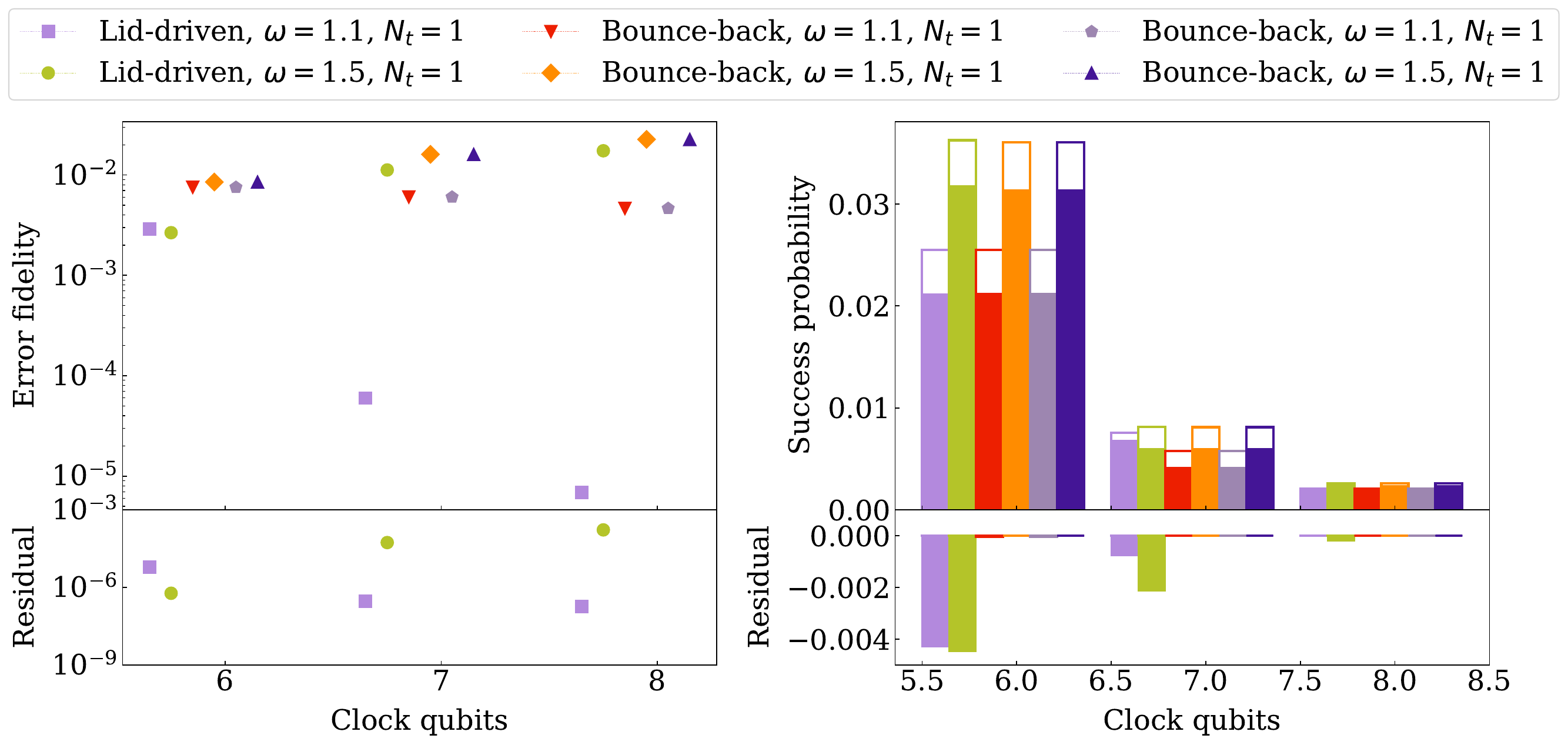}   \caption{Results obtained by approximating the spectrum of a larger lattice ($N_x=20$) using that of a smaller lattice ($N_x=4$) for single time-step evolution. The top panels display the fidelity error (left) and success probability (right) as a function of the number of clock qubits. 
The bottom panels show the residuals between the HHL results obtained using the approximated spectrum and those computed with the full $N_x=20$ spectrum. 
In the right panels, empty bars represent the probability of measuring the ancilla qubit in the $\ket{1}$ state, while filled bars correspond to measuring both the ancilla in  $\ket{1}$ and the clock qubits in the $\ket{0...0}$ state. Any missing points in the left residual panels indicate values of zero.}
 \label{fig:different_spectra_T1}
\end{figure*}

Figure~\ref{fig:different_spectra_T1} shows the fidelity error and success probability results (left and right panels respectively) as a function of clock qubits for a $20\times20$ lattice evolved for a single time step across three different systems: PBC, bounce-back, and lid-driven cavity. As in previous analyses, the PBC and bounce-back cases are initialized at $t_0=0$, while the lid-driven cavity starts from $t_0=40$. 
The top panels show the raw results, whereas the bottom panels illustrate the residuals between these results and those obtained from the full spectrum of the $20\times20$ lattice. This residual analysis quantifies the error introduced by the spectral approximation. Different colors and symbols correspond to different simulated systems. Points not visible in the residual plots indicate negligible or zero differences. We observe that the error introduced by using a smaller reference lattice is minimal, with fidelity residuals below $10^{-4}$. The only found change is in the success probability. These results confirm that approximating the spectrum using a smaller lattice does not significantly impact accuracy. More importantly, it mitigates one of the primary computational bottlenecks of the HHL algorithm—eigenvalue evaluation—leading to reduced circuit depth and improved efficiency.

We extend this analysis to a multi-time-step evolution, examining the dynamics of a $12\times 12$ lattice under the same conditions $\omega=1.1,\,1.5$ for three different applications. The system is evolved for three time steps,  with PBC and bounce-back initialized at $t_0=0$ and the lid-driven cavity at $t_0=40$. 
Figure~\ref{fig:different_spectra_T1} summarizes the results as as a function of the number of clock qubits. The left panels display the average fidelity errors, while the right panels show the success probability. 
As before, the top panels contain the absolute values, and the bottom panels report the residuals between the HHL results obtained with the approximated $4\times 4)$ spectrum and those using the full $12\times12$ spectrum. Despite slight spectral differences, the error introduced by using a smaller reference lattice remains negligible, with only the success probability being affected. As found previously, the approximation of spectrum using a smaller lattice does not change the performance of the proposed method.

\begin{figure*}[t]
    \centering
    \includegraphics[width=1\linewidth]{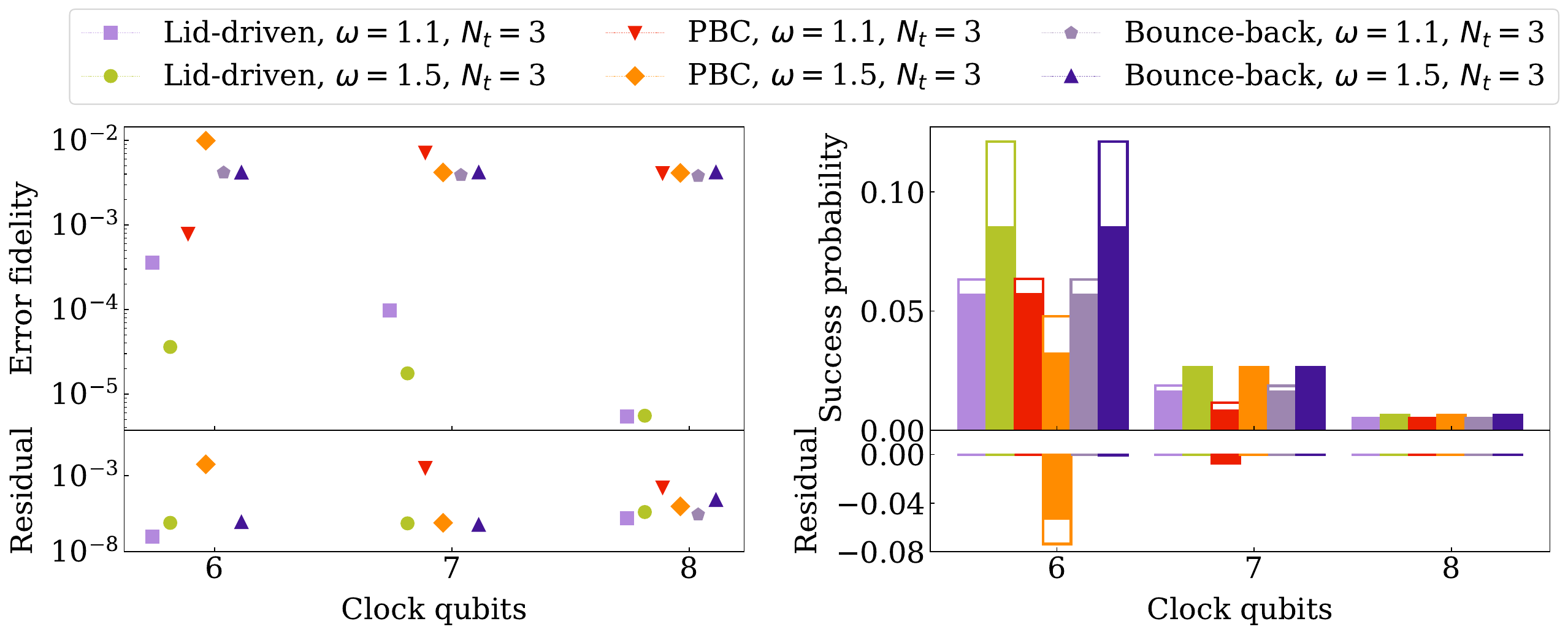}
   \caption{Results obtained by approximating the spectrum of a larger lattice ($N_x=12$) fusing that of a smaller lattice  ($N_x=4$) for three time-step evolution, starting from $t_0=40$.  The top panels display the fidelity error (left) and success probability (right) as a function of the number of clock qubits. The bottom panels show the residuals between the HHL results obtained using the approximated spectrum and those computed with the full $N_x=12$ spectrum. In the left panels, for each system and clock qubit configuration, we report the average fidelity error. In the right panels, empty bars represent the probability of measuring the ancilla qubit in the $\ket{1}$ state, while filled bars correspond to measuring both the ancilla in $\ket{1}$ and the clock qubits in the $\ket{0...0}$ state.}
\label{fig:different_spectra_T2}
\end{figure*}

\section{Conclusion\label{sec:conclusion}}

This work investigates a quantum Lattice Boltzmann method based on a Carleman linearization scheme combined with the application of the HHL algorithm, making use of exact state-vector emulation. 
While we primarily aim for long-term industrial applications, our investigation begins with simple 2D fluid systems: the lid-driven cavity, a closed box with bounce-back boundary conditions, and an open system with periodic boundary conditions (BC), all under various relaxation frequencies $\omega$.

Our analysis starts with evaluating the errors introduced by the Carleman linearization scheme's truncation procedure and proceeds to assess the performance of the HHL algorithm in solving the linearized LBM equations under certain assumptions. 
We find that the proposed hybrid method achieves satisfactory results, maintaining an error fidelity below $2\%$, even with a limited number of clock qubits, and demonstrates good success probabilities.
Notably, the quality, measured with the quantum fidelity, of the solution improves when evolving over multiple time steps, reaching an average error of $10^{-3}$ with an increase in success probability.

More importantly, we find that applying the HHL algorithm to large lattice sizes while utilizing spectra from smaller lattices yields high-quality results with minimal loss of accuracy. This demonstrates that approximating the spectra of larger systems using smaller lattice spectra can effectively reduce computational complexity without significantly compromising performance. Moreover, using this approximation combing the use of tensor network is under study to extend our simulations for larger lattice. 

An important challenge with the proposed method arises from the fact that the Carleman linearization scheme produces an error in approximating the LBM equations that is larger than that introduced by the use of HHL. 
The Carleman error is found to be approximately $5\%$, while the maximum error in HHL is under  $3\%$, at least for what concerns the exact state 
vector emulation. However, we expect that, increasing the Carleman truncation order, the approximation becomes more realistic. The cost is extremely high in classical computation, but from a quantum computing perspective, the number of qubits grows linearly with the Carleman order, facilitating the simulation of high turbulent systems. From the implemented HHL method, the precision of the solution is improved increasing the number of clock qubits at cost of a decreasing of success probability. Furthermore, solving Naiver Stokes equations with quantum computers can  be enhanced by implementing Carleman linearization scheme. 

Looking ahead, the ability to efficiently model fluid dynamics at the quantum level could pave the way for solving more intricate industrial problems that are currently computationally intensive on classical systems. Furthermore, as quantum computing hardware continues to improve,  the feasibility of applying quantum-enhanced methods to real-world fluid dynamics simulations will revolutionize the field, offering unprecedented accuracy and speed in modeling turbulence, heat transfer, and other complex phenomena. In the long term, this progress may lead to more optimized designs for aerodynamics, climate modeling, and material science, enabling industries to achieve breakthroughs that are presently constrained by the limits of classical computation.
The proposed method can help in this path.

\section{Acknowledgment}

The authors would like to express their gratitude to Matteo Zanfrognini for his patience in explaining the details of the Classical Lattice Boltzmann Method. They would also like to thank Claudio Sanavio for his valuable discussions on the Carleman LBM approach. Additionally, the authors extend their appreciation to Francesco Ferrari, Marco Maronese, members of the Leonardo S.p.A. Quantum Computing team, for their insightful contributions.

The authors acknowledge the financial support from ICSC-``National Research Centre in High Performance Computing, Big Data and Quantum Computing", funded by European Union-NextGenerationEU (CN00000013-Spoke 10).

%


\begin{thebibliography}{43}%
\makeatletter
\providecommand \@ifxundefined [1]{%
 \@ifx{#1\undefined}
}%
\providecommand \@ifnum [1]{%
 \ifnum #1\expandafter \@firstoftwo
 \else \expandafter \@secondoftwo
 \fi
}%
\providecommand \@ifx [1]{%
 \ifx #1\expandafter \@firstoftwo
 \else \expandafter \@secondoftwo
 \fi
}%
\providecommand \natexlab [1]{#1}%
\providecommand \enquote  [1]{``#1''}%
\providecommand \bibnamefont  [1]{#1}%
\providecommand \bibfnamefont [1]{#1}%
\providecommand \citenamefont [1]{#1}%
\providecommand \href@noop [0]{\@secondoftwo}%
\providecommand \href [0]{\begingroup \@sanitize@url \@href}%
\providecommand \@href[1]{\@@startlink{#1}\@@href}%
\providecommand \@@href[1]{\endgroup#1\@@endlink}%
\providecommand \@sanitize@url [0]{\catcode `\\12\catcode `\$12\catcode
  `\&12\catcode `\#12\catcode `\^12\catcode `\_12\catcode `\%12\relax}%
\providecommand \@@startlink[1]{}%
\providecommand \@@endlink[0]{}%
\providecommand \url  [0]{\begingroup\@sanitize@url \@url }%
\providecommand \@url [1]{\endgroup\@href {#1}{\urlprefix }}%
\providecommand \urlprefix  [0]{URL }%
\providecommand \Eprint [0]{\href }%
\providecommand \doibase [0]{https://doi.org/}%
\providecommand \selectlanguage [0]{\@gobble}%
\providecommand \bibinfo  [0]{\@secondoftwo}%
\providecommand \bibfield  [0]{\@secondoftwo}%
\providecommand \translation [1]{[#1]}%
\providecommand \BibitemOpen [0]{}%
\providecommand \bibitemStop [0]{}%
\providecommand \bibitemNoStop [0]{.\EOS\space}%
\providecommand \EOS [0]{\spacefactor3000\relax}%
\providecommand \BibitemShut  [1]{\csname bibitem#1\endcsname}%
\let\auto@bib@innerbib\@empty
\bibitem [{\citenamefont {McLean}(2012)}]{McLean2012}%
  \BibitemOpen
  \bibfield  {author} {\bibinfo {author} {\bibfnamefont {D.}~\bibnamefont
  {McLean}},\ }\href {https://doi.org/10.1002/9781118454190} {\emph {\bibinfo
  {title} {Understanding Aerodynamics: Arguing from the Real Physics}}}\
  (\bibinfo  {publisher} {Wiley},\ \bibinfo {year} {2012})\BibitemShut
  {NoStop}%
\bibitem [{\citenamefont {Norton}\ \emph {et~al.}(2007)\citenamefont {Norton},
  \citenamefont {Sun}, \citenamefont {Grant}, \citenamefont {Fallon},\ and\
  \citenamefont {Dodd}}]{industrial_application_1}%
  \BibitemOpen
  \bibfield  {author} {\bibinfo {author} {\bibfnamefont {T.}~\bibnamefont
  {Norton}}, \bibinfo {author} {\bibfnamefont {D.-W.}\ \bibnamefont {Sun}},
  \bibinfo {author} {\bibfnamefont {J.}~\bibnamefont {Grant}}, \bibinfo
  {author} {\bibfnamefont {R.}~\bibnamefont {Fallon}},\ and\ \bibinfo {author}
  {\bibfnamefont {V.}~\bibnamefont {Dodd}},\ }\bibfield  {title} {\bibinfo
  {title} {Applications of computational fluid dynamics (cfd) in the modelling
  and design of ventilation systems in the agricultural industry: A review},\
  }\href {https://doi.org/10.1016/j.biortech.2006.11.025} {\bibfield  {journal}
  {\bibinfo  {journal} {Bioresource Technology}\ }\textbf {\bibinfo {volume}
  {98}},\ \bibinfo {pages} {2386–2414} (\bibinfo {year} {2007})}\BibitemShut
  {NoStop}%
\bibitem [{\citenamefont {Mani}\ and\ \citenamefont
  {Dorgan}(2023)}]{industrial_application_2}%
  \BibitemOpen
  \bibfield  {author} {\bibinfo {author} {\bibfnamefont {M.}~\bibnamefont
  {Mani}}\ and\ \bibinfo {author} {\bibfnamefont {A.~J.}\ \bibnamefont
  {Dorgan}},\ }\bibfield  {title} {\bibinfo {title} {A perspective on the state
  of aerospace computational fluid dynamics technology},\ }\href
  {https://doi.org/10.1146/annurev-fluid-120720-124800} {\bibfield  {journal}
  {\bibinfo  {journal} {Annual Review of Fluid Mechanics}\ }\textbf {\bibinfo
  {volume} {55}},\ \bibinfo {pages} {431–457} (\bibinfo {year}
  {2023})}\BibitemShut {NoStop}%
\bibitem [{\citenamefont {Ferro}\ \emph {et~al.}(2023)\citenamefont {Ferro},
  \citenamefont {Maggiore},\ and\ \citenamefont
  {Champvillair}}]{industrial_application_3}%
  \BibitemOpen
  \bibfield  {author} {\bibinfo {author} {\bibfnamefont {C.~G.}\ \bibnamefont
  {Ferro}}, \bibinfo {author} {\bibfnamefont {P.}~\bibnamefont {Maggiore}},\
  and\ \bibinfo {author} {\bibfnamefont {D.}~\bibnamefont {Champvillair}},\
  }\bibfield  {title} {\bibinfo {title} {Development of a computational fluid
  dynamics model for ice formation: Validation and parameter analysis},\ }\href
  {https://doi.org/10.3390/atmos14050834} {\bibfield  {journal} {\bibinfo
  {journal} {Atmosphere}\ }\textbf {\bibinfo {volume} {14}},\ \bibinfo {pages}
  {834} (\bibinfo {year} {2023})}\BibitemShut {NoStop}%
\bibitem [{\citenamefont {Zajaczkowski}\ \emph {et~al.}(2011)\citenamefont
  {Zajaczkowski}, \citenamefont {Haupt},\ and\ \citenamefont
  {Schmehl}}]{industrial_application_4}%
  \BibitemOpen
  \bibfield  {author} {\bibinfo {author} {\bibfnamefont {F.~J.}\ \bibnamefont
  {Zajaczkowski}}, \bibinfo {author} {\bibfnamefont {S.~E.}\ \bibnamefont
  {Haupt}},\ and\ \bibinfo {author} {\bibfnamefont {K.~J.}\ \bibnamefont
  {Schmehl}},\ }\bibfield  {title} {\bibinfo {title} {A preliminary study of
  assimilating numerical weather prediction data into computational fluid
  dynamics models for wind prediction},\ }\href
  {https://doi.org/10.1016/j.jweia.2011.01.023} {\bibfield  {journal} {\bibinfo
   {journal} {Journal of Wind Engineering and Industrial Aerodynamics}\
  }\textbf {\bibinfo {volume} {99}},\ \bibinfo {pages} {320–329} (\bibinfo
  {year} {2011})}\BibitemShut {NoStop}%
\bibitem [{\citenamefont {Jujjavarapu}\ \emph {et~al.}(2024)\citenamefont
  {Jujjavarapu}, \citenamefont {Kumar},\ and\ \citenamefont
  {Gupta}}]{industrial_application_5}%
  \BibitemOpen
  \bibfield  {author} {\bibinfo {author} {\bibfnamefont {S.~E.}\ \bibnamefont
  {Jujjavarapu}}, \bibinfo {author} {\bibfnamefont {T.}~\bibnamefont {Kumar}},\
  and\ \bibinfo {author} {\bibfnamefont {S.}~\bibnamefont {Gupta}},\ }\bibinfo
  {title} {Computational fluid dynamics in biomedical engineering},\ in\ \href
  {https://doi.org/10.1007/978-981-99-7129-9_4} {\emph {\bibinfo {booktitle}
  {Computational Fluid Dynamics Applications in Bio and Biomedical
  Processes}}}\ (\bibinfo  {publisher} {Springer Nature Singapore},\ \bibinfo
  {year} {2024})\ p.\ \bibinfo {pages} {101–125}\BibitemShut {NoStop}%
\bibitem [{\citenamefont {Pope}(2000)}]{Pope2000}%
  \BibitemOpen
  \bibfield  {author} {\bibinfo {author} {\bibfnamefont {S.~B.}\ \bibnamefont
  {Pope}},\ }\href {https://doi.org/10.1017/cbo9780511840531} {\emph {\bibinfo
  {title} {Turbulent Flows}}}\ (\bibinfo  {publisher} {Cambridge University
  Press},\ \bibinfo {year} {2000})\BibitemShut {NoStop}%
\bibitem [{\citenamefont {Schlichting}\ and\ \citenamefont
  {Gersten}(2000)}]{Schlichting2000}%
  \BibitemOpen
  \bibfield  {author} {\bibinfo {author} {\bibfnamefont {H.}~\bibnamefont
  {Schlichting}}\ and\ \bibinfo {author} {\bibfnamefont {K.}~\bibnamefont
  {Gersten}},\ }\href {https://doi.org/10.1007/978-3-642-85829-1} {\emph
  {\bibinfo {title} {Boundary-Layer Theory}}}\ (\bibinfo  {publisher} {Springer
  Berlin Heidelberg},\ \bibinfo {year} {2000})\BibitemShut {NoStop}%
\bibitem [{\citenamefont {Gaitan}(2020)}]{gaitan2020finding}%
  \BibitemOpen
  \bibfield  {author} {\bibinfo {author} {\bibfnamefont {F.}~\bibnamefont
  {Gaitan}},\ }\bibfield  {title} {\bibinfo {title} {Finding flows of a
  navier--stokes fluid through quantum computing},\ }\href
  {https://doi.org/10.1038/s41534-020-00291-0} {\bibfield  {journal} {\bibinfo
  {journal} {npj Quantum Inf}\ }\textbf {\bibinfo {volume} {6}},\ \bibinfo
  {pages} {61} (\bibinfo {year} {2020})}\BibitemShut {NoStop}%
\bibitem [{\citenamefont {Meng}\ and\ \citenamefont
  {Yang}(2023)}]{PhysRevResearch.5.033182}%
  \BibitemOpen
  \bibfield  {author} {\bibinfo {author} {\bibfnamefont {Z.}~\bibnamefont
  {Meng}}\ and\ \bibinfo {author} {\bibfnamefont {Y.}~\bibnamefont {Yang}},\
  }\bibfield  {title} {\bibinfo {title} {Quantum computing of fluid dynamics
  using the hydrodynamic schr\"odinger equation},\ }\href
  {https://doi.org/10.1103/PhysRevResearch.5.033182} {\bibfield  {journal}
  {\bibinfo  {journal} {Phys. Rev. Res.}\ }\textbf {\bibinfo {volume} {5}},\
  \bibinfo {pages} {033182} (\bibinfo {year} {2023})}\BibitemShut {NoStop}%
\bibitem [{\citenamefont {Meng}\ \emph {et~al.}(2024)\citenamefont {Meng},
  \citenamefont {Zhong}, \citenamefont {Xu}, \citenamefont {Wang},
  \citenamefont {Chen}, \citenamefont {Jin}, \citenamefont {Zhu}, \citenamefont
  {Gao}, \citenamefont {Wu}, \citenamefont {Zhang} \emph
  {et~al.}}]{meng2024simulating}%
  \BibitemOpen
  \bibfield  {author} {\bibinfo {author} {\bibfnamefont {Z.}~\bibnamefont
  {Meng}}, \bibinfo {author} {\bibfnamefont {J.}~\bibnamefont {Zhong}},
  \bibinfo {author} {\bibfnamefont {S.}~\bibnamefont {Xu}}, \bibinfo {author}
  {\bibfnamefont {K.}~\bibnamefont {Wang}}, \bibinfo {author} {\bibfnamefont
  {J.}~\bibnamefont {Chen}}, \bibinfo {author} {\bibfnamefont {F.}~\bibnamefont
  {Jin}}, \bibinfo {author} {\bibfnamefont {X.}~\bibnamefont {Zhu}}, \bibinfo
  {author} {\bibfnamefont {Y.}~\bibnamefont {Gao}}, \bibinfo {author}
  {\bibfnamefont {Y.}~\bibnamefont {Wu}}, \bibinfo {author} {\bibfnamefont
  {C.}~\bibnamefont {Zhang}}, \emph {et~al.},\ }\bibfield  {title} {\bibinfo
  {title} {Simulating unsteady flows on a superconducting quantum processor},\
  }\href {https://doi.org/10.1038/s42005-024-01845-w} {\bibfield  {journal}
  {\bibinfo  {journal} {Communications Physics}\ }\textbf {\bibinfo {volume}
  {7}},\ \bibinfo {pages} {349} (\bibinfo {year} {2024})}\BibitemShut {NoStop}%
\bibitem [{\citenamefont {Chen}\ \emph {et~al.}(2022)\citenamefont {Chen},
  \citenamefont {Xue}, \citenamefont {Chen}, \citenamefont {Lu}, \citenamefont
  {Wu}, \citenamefont {Ding}, \citenamefont {Huang},\ and\ \citenamefont
  {Guo}}]{chen2022quantumfem}%
  \BibitemOpen
  \bibfield  {author} {\bibinfo {author} {\bibfnamefont {Z.-Y.}\ \bibnamefont
  {Chen}}, \bibinfo {author} {\bibfnamefont {C.}~\bibnamefont {Xue}}, \bibinfo
  {author} {\bibfnamefont {S.-M.}\ \bibnamefont {Chen}}, \bibinfo {author}
  {\bibfnamefont {B.-H.}\ \bibnamefont {Lu}}, \bibinfo {author} {\bibfnamefont
  {Y.-C.}\ \bibnamefont {Wu}}, \bibinfo {author} {\bibfnamefont {J.-C.}\
  \bibnamefont {Ding}}, \bibinfo {author} {\bibfnamefont {S.-H.}\ \bibnamefont
  {Huang}},\ and\ \bibinfo {author} {\bibfnamefont {G.-P.}\ \bibnamefont
  {Guo}},\ }\bibfield  {title} {\bibinfo {title} {Quantum approach to
  accelerate finite volume method on steady computational fluid dynamics
  problems},\ }\href {https://doi.org/10.1007/s11128-022-03478-w} {\bibfield
  {journal} {\bibinfo  {journal} {Quantum Information Processing}\ }\textbf
  {\bibinfo {volume} {21}},\ \bibinfo {pages} {137} (\bibinfo {year}
  {2022})}\BibitemShut {NoStop}%
\bibitem [{\citenamefont {Oz}\ \emph {et~al.}(2022)\citenamefont {Oz},
  \citenamefont {Vuppala}, \citenamefont {Kara},\ and\ \citenamefont
  {Gaitan}}]{oz2022solving}%
  \BibitemOpen
  \bibfield  {author} {\bibinfo {author} {\bibfnamefont {F.}~\bibnamefont
  {Oz}}, \bibinfo {author} {\bibfnamefont {R.~K.}\ \bibnamefont {Vuppala}},
  \bibinfo {author} {\bibfnamefont {K.}~\bibnamefont {Kara}},\ and\ \bibinfo
  {author} {\bibfnamefont {F.}~\bibnamefont {Gaitan}},\ }\bibfield  {title}
  {\bibinfo {title} {Solving burgers’ equation with quantum computing},\
  }\href {https://doi.org/10.1007/s11128-021-03391-8} {\bibfield  {journal}
  {\bibinfo  {journal} {Quantum Inf Process}\ }\textbf {\bibinfo {volume}
  {21}},\ \bibinfo {pages} {30} (\bibinfo {year} {2022})}\BibitemShut {NoStop}%
\bibitem [{\citenamefont {Liu}\ \emph {et~al.}(2021)\citenamefont {Liu},
  \citenamefont {Kolden}, \citenamefont {Krovi}, \citenamefont {Loureiro},
  \citenamefont {Trivisa},\ and\ \citenamefont {Childs}}]{liu2021efficient}%
  \BibitemOpen
  \bibfield  {author} {\bibinfo {author} {\bibfnamefont {J.-P.}\ \bibnamefont
  {Liu}}, \bibinfo {author} {\bibfnamefont {H.~{\O}.}\ \bibnamefont {Kolden}},
  \bibinfo {author} {\bibfnamefont {H.~K.}\ \bibnamefont {Krovi}}, \bibinfo
  {author} {\bibfnamefont {N.~F.}\ \bibnamefont {Loureiro}}, \bibinfo {author}
  {\bibfnamefont {K.}~\bibnamefont {Trivisa}},\ and\ \bibinfo {author}
  {\bibfnamefont {A.~M.}\ \bibnamefont {Childs}},\ }\bibfield  {title}
  {\bibinfo {title} {Efficient quantum algorithm for dissipative nonlinear
  differential equations},\ }\href {https://doi.org/10.1073/pnas.2026805118}
  {\bibfield  {journal} {\bibinfo  {journal} {Proc Natl Acad Scis}\ }\textbf
  {\bibinfo {volume} {118}},\ \bibinfo {pages} {e2026805118} (\bibinfo {year}
  {2021})}\BibitemShut {NoStop}%
\bibitem [{\citenamefont {Gaitan}(2021)}]{gaitan2021finding}%
  \BibitemOpen
  \bibfield  {author} {\bibinfo {author} {\bibfnamefont {F.}~\bibnamefont
  {Gaitan}},\ }\bibfield  {title} {\bibinfo {title} {Finding solutions of the
  navier-stokes equations through quantum computing—recent progress, a
  generalization, and next steps forward},\ }\href
  {https://doi.org/10.1002/qute.202100055} {\bibfield  {journal} {\bibinfo
  {journal} {Adv. Quantum Technol}\ }\textbf {\bibinfo {volume} {4}},\ \bibinfo
  {pages} {2100055} (\bibinfo {year} {2021})}\BibitemShut {NoStop}%
\bibitem [{\citenamefont {Li}\ \emph {et~al.}(2025)\citenamefont {Li},
  \citenamefont {Yin}, \citenamefont {Wiebe}, \citenamefont {Chun},
  \citenamefont {Schenter}, \citenamefont {Cheung},\ and\ \citenamefont
  {M\"ulmenst\"adt}}]{li2024potentialquantumadvantagesimulation}%
  \BibitemOpen
  \bibfield  {author} {\bibinfo {author} {\bibfnamefont {X.}~\bibnamefont
  {Li}}, \bibinfo {author} {\bibfnamefont {X.}~\bibnamefont {Yin}}, \bibinfo
  {author} {\bibfnamefont {N.}~\bibnamefont {Wiebe}}, \bibinfo {author}
  {\bibfnamefont {J.}~\bibnamefont {Chun}}, \bibinfo {author} {\bibfnamefont
  {G.~K.}\ \bibnamefont {Schenter}}, \bibinfo {author} {\bibfnamefont {M.~S.}\
  \bibnamefont {Cheung}},\ and\ \bibinfo {author} {\bibfnamefont
  {J.}~\bibnamefont {M\"ulmenst\"adt}},\ }\href
  {https://doi.org/10.1103/PhysRevResearch.7.013036} {\bibinfo {title}
  {Potential quantum advantage for simulation of fluid dynamics}} (\bibinfo
  {year} {2025})\BibitemShut {NoStop}%
\bibitem [{\citenamefont {Montanaro}\ and\ \citenamefont
  {Pallister}(2016)}]{Montanaro2016fem}%
  \BibitemOpen
  \bibfield  {author} {\bibinfo {author} {\bibfnamefont {A.}~\bibnamefont
  {Montanaro}}\ and\ \bibinfo {author} {\bibfnamefont {S.}~\bibnamefont
  {Pallister}},\ }\bibfield  {title} {\bibinfo {title} {Quantum algorithms and
  the finite element method},\ }\href
  {https://doi.org/10.1103/PhysRevA.93.032324} {\bibfield  {journal} {\bibinfo
  {journal} {Phys. Rev. A}\ }\textbf {\bibinfo {volume} {93}},\ \bibinfo
  {pages} {032324} (\bibinfo {year} {2016})}\BibitemShut {NoStop}%
\bibitem [{\citenamefont {Jin}\ \emph {et~al.}(2022)\citenamefont {Jin},
  \citenamefont {Liu},\ and\ \citenamefont {Yu}}]{jin2022time}%
  \BibitemOpen
  \bibfield  {author} {\bibinfo {author} {\bibfnamefont {S.}~\bibnamefont
  {Jin}}, \bibinfo {author} {\bibfnamefont {N.}~\bibnamefont {Liu}},\ and\
  \bibinfo {author} {\bibfnamefont {Y.}~\bibnamefont {Yu}},\ }\bibfield
  {title} {\bibinfo {title} {Time complexity analysis of quantum difference
  methods for linear high dimensional and multiscale partial differential
  equations},\ }\href
  {https://doi.org/https://doi.org/10.1016/j.jcp.2022.111641} {\bibfield
  {journal} {\bibinfo  {journal} {Journal of Computational Physics}\ }\textbf
  {\bibinfo {volume} {471}},\ \bibinfo {pages} {111641} (\bibinfo {year}
  {2022})}\BibitemShut {NoStop}%
\bibitem [{\citenamefont {Succi}(2001)}]{succi2001lattice}%
  \BibitemOpen
  \bibfield  {author} {\bibinfo {author} {\bibfnamefont {S.}~\bibnamefont
  {Succi}},\ }\href@noop {} {\emph {\bibinfo {title} {The Lattice Boltzmann
  Equation for Fluid Dynamics and Beyond}}}\ (\bibinfo  {publisher} {Oxford
  University Press},\ \bibinfo {year} {2001})\BibitemShut {NoStop}%
\bibitem [{\citenamefont {Chen}\ and\ \citenamefont {Doolen}(1998)}]{Chen1998}%
  \BibitemOpen
  \bibfield  {author} {\bibinfo {author} {\bibfnamefont {S.}~\bibnamefont
  {Chen}}\ and\ \bibinfo {author} {\bibfnamefont {G.~D.}\ \bibnamefont
  {Doolen}},\ }\bibfield  {title} {\bibinfo {title} {Lattice boltzmann method
  for fluid flows},\ }\href {https://doi.org/10.1146/annurev.fluid.30.1.329}
  {\bibfield  {journal} {\bibinfo  {journal} {Annual Review of Fluid
  Mechanics}\ }\textbf {\bibinfo {volume} {30}},\ \bibinfo {pages} {329–364}
  (\bibinfo {year} {1998})}\BibitemShut {NoStop}%
\bibitem [{\citenamefont {Itani}\ \emph
  {et~al.}(2023{\natexlab{a}})\citenamefont {Itani}, \citenamefont
  {Sreenivasan},\ and\ \citenamefont {Succi}}]{itani2023}%
  \BibitemOpen
  \bibfield  {author} {\bibinfo {author} {\bibfnamefont {W.}~\bibnamefont
  {Itani}}, \bibinfo {author} {\bibfnamefont {K.~R.}\ \bibnamefont
  {Sreenivasan}},\ and\ \bibinfo {author} {\bibfnamefont {S.}~\bibnamefont
  {Succi}},\ }\href {https://arxiv.org/abs/2301.05762} {\bibinfo {title}
  {Quantum carleman lattice boltzmann simulation of fluids}} (\bibinfo {year}
  {2023}{\natexlab{a}}),\ \Eprint {https://arxiv.org/abs/2301.05762}
  {arXiv:2301.05762 [physics.flu-dyn]} \BibitemShut {NoStop}%
\bibitem [{\citenamefont {Itani}\ \emph
  {et~al.}(2023{\natexlab{b}})\citenamefont {Itani}, \citenamefont
  {Sreenivasan},\ and\ \citenamefont
  {Succi}}]{itani2023quantumalgorithmlatticeboltzmann}%
  \BibitemOpen
  \bibfield  {author} {\bibinfo {author} {\bibfnamefont {W.}~\bibnamefont
  {Itani}}, \bibinfo {author} {\bibfnamefont {K.~R.}\ \bibnamefont
  {Sreenivasan}},\ and\ \bibinfo {author} {\bibfnamefont {S.}~\bibnamefont
  {Succi}},\ }\href {https://arxiv.org/abs/2304.05915} {\bibinfo {title}
  {Quantum algorithm for lattice boltzmann (qalb) simulation of incompressible
  fluids with a nonlinear collision term}} (\bibinfo {year}
  {2023}{\natexlab{b}}),\ \Eprint {https://arxiv.org/abs/2304.05915}
  {arXiv:2304.05915 [quant-ph]} \BibitemShut {NoStop}%
\bibitem [{\citenamefont {Kocherla}\ \emph {et~al.}(2024)\citenamefont
  {Kocherla}, \citenamefont {Song}, \citenamefont {Chrit}, \citenamefont
  {Gard}, \citenamefont {Dumitrescu}, \citenamefont {Alexeev},\ and\
  \citenamefont {Bryngelson}}]{Kocherla2024Fullquantumalgorithm}%
  \BibitemOpen
  \bibfield  {author} {\bibinfo {author} {\bibfnamefont {S.}~\bibnamefont
  {Kocherla}}, \bibinfo {author} {\bibfnamefont {Z.}~\bibnamefont {Song}},
  \bibinfo {author} {\bibfnamefont {F.~E.}\ \bibnamefont {Chrit}}, \bibinfo
  {author} {\bibfnamefont {B.}~\bibnamefont {Gard}}, \bibinfo {author}
  {\bibfnamefont {E.~F.}\ \bibnamefont {Dumitrescu}}, \bibinfo {author}
  {\bibfnamefont {A.}~\bibnamefont {Alexeev}},\ and\ \bibinfo {author}
  {\bibfnamefont {S.~H.}\ \bibnamefont {Bryngelson}},\ }\bibfield  {title}
  {\bibinfo {title} {Fully quantum algorithm for mesoscale fluid simulations
  with application to partial differential equations},\ }\href
  {https://doi.org/10.1116/5.0217675} {\bibfield  {journal} {\bibinfo
  {journal} {AVS Quantum Science}\ }\textbf {\bibinfo {volume} {6}},\ \bibinfo
  {pages} {033806} (\bibinfo {year} {2024})}\BibitemShut {NoStop}%
\bibitem [{\citenamefont {Schalkers}\ and\ \citenamefont
  {M\"{o}ller}(2024)}]{10.1016/j.jcp.2024.112816}%
  \BibitemOpen
  \bibfield  {author} {\bibinfo {author} {\bibfnamefont {M.~A.}\ \bibnamefont
  {Schalkers}}\ and\ \bibinfo {author} {\bibfnamefont {M.}~\bibnamefont
  {M\"{o}ller}},\ }\bibfield  {title} {\bibinfo {title} {Efficient and
  fail-safe quantum algorithm for the transport equation},\ }\bibfield
  {journal} {\bibinfo  {journal} {J. Comput. Phys.}\ }\textbf {\bibinfo
  {volume} {502}},\ \href {https://doi.org/10.1016/j.jcp.2024.112816}
  {10.1016/j.jcp.2024.112816} (\bibinfo {year} {2024})\BibitemShut {NoStop}%
\bibitem [{\citenamefont {Schalkers}\ and\ \citenamefont
  {M{\"o}ller}(2024)}]{schalkers2024importance}%
  \BibitemOpen
  \bibfield  {author} {\bibinfo {author} {\bibfnamefont {M.~A.}\ \bibnamefont
  {Schalkers}}\ and\ \bibinfo {author} {\bibfnamefont {M.}~\bibnamefont
  {M{\"o}ller}},\ }\bibfield  {title} {\bibinfo {title} {On the importance of
  data encoding in quantum boltzmann methods},\ }\href
  {https://doi.org/10.1007/s11128-023-04216-6} {\bibfield  {journal} {\bibinfo
  {journal} {Quantum Inf Process}\ }\textbf {\bibinfo {volume} {23}},\ \bibinfo
  {pages} {20} (\bibinfo {year} {2024})}\BibitemShut {NoStop}%
\bibitem [{\citenamefont {Georgescu}\ \emph {et~al.}(2024)\citenamefont
  {Georgescu}, \citenamefont {Schalkers},\ and\ \citenamefont
  {Möller}}]{georgescu2024qlbmquantumlattice}%
  \BibitemOpen
  \bibfield  {author} {\bibinfo {author} {\bibfnamefont {C.~A.}\ \bibnamefont
  {Georgescu}}, \bibinfo {author} {\bibfnamefont {M.~A.}\ \bibnamefont
  {Schalkers}},\ and\ \bibinfo {author} {\bibfnamefont {M.}~\bibnamefont
  {Möller}},\ }\href {https://arxiv.org/abs/2411.19439} {\bibinfo {title}
  {qlbm -- a quantum lattice boltzmann software framework}} (\bibinfo {year}
  {2024}),\ \Eprint {https://arxiv.org/abs/2411.19439} {arXiv:2411.19439
  [quant-ph]} \BibitemShut {NoStop}%
\bibitem [{\citenamefont {Carleman}(1932)}]{Carleman1932}%
  \BibitemOpen
  \bibfield  {author} {\bibinfo {author} {\bibfnamefont {T.}~\bibnamefont
  {Carleman}},\ }\bibfield  {title} {\bibinfo {title} {Application de la
  théorie des équations intégrales linéaires aux systèmes d’équations
  différentielles non linéaires},\ }\href
  {https://doi.org/10.1007/bf02546499} {\bibfield  {journal} {\bibinfo
  {journal} {Acta Mathematica}\ }\textbf {\bibinfo {volume} {59}},\ \bibinfo
  {pages} {63–87} (\bibinfo {year} {1932})}\BibitemShut {NoStop}%
\bibitem [{\citenamefont {Sanavio}\ \emph {et~al.}(2024)\citenamefont
  {Sanavio}, \citenamefont {Mauri},\ and\ \citenamefont
  {Succi}}]{sanavio2024gradapproach}%
  \BibitemOpen
  \bibfield  {author} {\bibinfo {author} {\bibfnamefont {C.}~\bibnamefont
  {Sanavio}}, \bibinfo {author} {\bibfnamefont {E.}~\bibnamefont {Mauri}},\
  and\ \bibinfo {author} {\bibfnamefont {S.}~\bibnamefont {Succi}},\ }\href
  {https://arxiv.org/abs/2406.01118} {\bibinfo {title} {Carleman-grad approach
  to the quantum simulation of fluids}} (\bibinfo {year} {2024}),\ \Eprint
  {https://arxiv.org/abs/2406.01118} {arXiv:2406.01118 [quant-ph]} \BibitemShut
  {NoStop}%
\bibitem [{\citenamefont {Sanavio}\ \emph {et~al.}(2025)\citenamefont
  {Sanavio}, \citenamefont {Simon}, \citenamefont {Ralli}, \citenamefont
  {Love},\ and\ \citenamefont {Succi}}]{sanavio2025carleman}%
  \BibitemOpen
  \bibfield  {author} {\bibinfo {author} {\bibfnamefont {C.}~\bibnamefont
  {Sanavio}}, \bibinfo {author} {\bibfnamefont {W.~A.}\ \bibnamefont {Simon}},
  \bibinfo {author} {\bibfnamefont {A.}~\bibnamefont {Ralli}}, \bibinfo
  {author} {\bibfnamefont {P.}~\bibnamefont {Love}},\ and\ \bibinfo {author}
  {\bibfnamefont {S.}~\bibnamefont {Succi}},\ }\href
  {https://arxiv.org/abs/2501.02582} {\bibinfo {title}
  {Carleman-lattice-boltzmann quantum circuit with matrix access oracles}}
  (\bibinfo {year} {2025}),\ \Eprint {https://arxiv.org/abs/2501.02582}
  {arXiv:2501.02582 [quant-ph]} \BibitemShut {NoStop}%
\bibitem [{\citenamefont {Zaman}\ \emph {et~al.}(2023)\citenamefont {Zaman},
  \citenamefont {Morrell},\ and\ \citenamefont {Wong}}]{zaman2023step}%
  \BibitemOpen
  \bibfield  {author} {\bibinfo {author} {\bibfnamefont {A.}~\bibnamefont
  {Zaman}}, \bibinfo {author} {\bibfnamefont {H.~J.}\ \bibnamefont {Morrell}},\
  and\ \bibinfo {author} {\bibfnamefont {H.~Y.}\ \bibnamefont {Wong}},\
  }\bibfield  {title} {\bibinfo {title} {A step-by-step hhl algorithm
  walkthrough to enhance understanding of critical quantum computing
  concepts},\ }\bibfield  {journal} {\bibinfo  {journal} {IEEE Access}\ }\href
  {https://doi.org/10.1109/ACCESS.2023.3297658} {10.1109/ACCESS.2023.3297658}
  (\bibinfo {year} {2023})\BibitemShut {NoStop}%
\bibitem [{\citenamefont {Cai}\ \emph {et~al.}(2013)\citenamefont {Cai},
  \citenamefont {Weedbrook}, \citenamefont {Su}, \citenamefont {Chen},
  \citenamefont {Gu}, \citenamefont {Zhu}, \citenamefont {Li}, \citenamefont
  {Liu}, \citenamefont {Lu},\ and\ \citenamefont
  {Pan}}]{PhysRevLett.110.230501}%
  \BibitemOpen
  \bibfield  {author} {\bibinfo {author} {\bibfnamefont {X.-D.}\ \bibnamefont
  {Cai}}, \bibinfo {author} {\bibfnamefont {C.}~\bibnamefont {Weedbrook}},
  \bibinfo {author} {\bibfnamefont {Z.-E.}\ \bibnamefont {Su}}, \bibinfo
  {author} {\bibfnamefont {M.-C.}\ \bibnamefont {Chen}}, \bibinfo {author}
  {\bibfnamefont {M.}~\bibnamefont {Gu}}, \bibinfo {author} {\bibfnamefont
  {M.-J.}\ \bibnamefont {Zhu}}, \bibinfo {author} {\bibfnamefont
  {L.}~\bibnamefont {Li}}, \bibinfo {author} {\bibfnamefont {N.-L.}\
  \bibnamefont {Liu}}, \bibinfo {author} {\bibfnamefont {C.-Y.}\ \bibnamefont
  {Lu}},\ and\ \bibinfo {author} {\bibfnamefont {J.-W.}\ \bibnamefont {Pan}},\
  }\bibfield  {title} {\bibinfo {title} {Experimental quantum computing to
  solve systems of linear equations},\ }\href
  {https://doi.org/10.1103/PhysRevLett.110.230501} {\bibfield  {journal}
  {\bibinfo  {journal} {Phys. Rev. Lett.}\ }\textbf {\bibinfo {volume} {110}},\
  \bibinfo {pages} {230501} (\bibinfo {year} {2013})}\BibitemShut {NoStop}%
\bibitem [{\citenamefont {Dervovic}\ \emph {et~al.}(2018)\citenamefont
  {Dervovic}, \citenamefont {Herbster}, \citenamefont {Mountney}, \citenamefont
  {Severini}, \citenamefont {Usher},\ and\ \citenamefont
  {Wossnig}}]{dervovic2018quantumlinearsystemsalgorithms}%
  \BibitemOpen
  \bibfield  {author} {\bibinfo {author} {\bibfnamefont {D.}~\bibnamefont
  {Dervovic}}, \bibinfo {author} {\bibfnamefont {M.}~\bibnamefont {Herbster}},
  \bibinfo {author} {\bibfnamefont {P.}~\bibnamefont {Mountney}}, \bibinfo
  {author} {\bibfnamefont {S.}~\bibnamefont {Severini}}, \bibinfo {author}
  {\bibfnamefont {N.}~\bibnamefont {Usher}},\ and\ \bibinfo {author}
  {\bibfnamefont {L.}~\bibnamefont {Wossnig}},\ }\href
  {https://arxiv.org/abs/1802.08227} {\bibinfo {title} {Quantum linear systems
  algorithms: a primer}} (\bibinfo {year} {2018}),\ \Eprint
  {https://arxiv.org/abs/1802.08227} {arXiv:1802.08227 [quant-ph]} \BibitemShut
  {NoStop}%
\bibitem [{\citenamefont {Harrow}\ \emph {et~al.}(2009)\citenamefont {Harrow},
  \citenamefont {Hassidim},\ and\ \citenamefont {Lloyd}}]{hhl}%
  \BibitemOpen
  \bibfield  {author} {\bibinfo {author} {\bibfnamefont {A.~W.}\ \bibnamefont
  {Harrow}}, \bibinfo {author} {\bibfnamefont {A.}~\bibnamefont {Hassidim}},\
  and\ \bibinfo {author} {\bibfnamefont {S.}~\bibnamefont {Lloyd}},\ }\bibfield
   {title} {\bibinfo {title} {Quantum algorithm for linear systems of
  equations},\ }\bibfield  {journal} {\bibinfo  {journal} {Physical Review
  Letters}\ }\textbf {\bibinfo {volume} {103}},\ \href
  {https://doi.org/10.1103/physrevlett.103.150502}
  {10.1103/physrevlett.103.150502} (\bibinfo {year} {2009})\BibitemShut
  {NoStop}%
\bibitem [{\citenamefont {Bravo-Prieto}\ \emph {et~al.}(2023)\citenamefont
  {Bravo-Prieto}, \citenamefont {LaRose}, \citenamefont {Cerezo}, \citenamefont
  {Subasi}, \citenamefont {Cincio},\ and\ \citenamefont
  {Coles}}]{BravoPrieto2023variationalquantum}%
  \BibitemOpen
  \bibfield  {author} {\bibinfo {author} {\bibfnamefont {C.}~\bibnamefont
  {Bravo-Prieto}}, \bibinfo {author} {\bibfnamefont {R.}~\bibnamefont
  {LaRose}}, \bibinfo {author} {\bibfnamefont {M.}~\bibnamefont {Cerezo}},
  \bibinfo {author} {\bibfnamefont {Y.}~\bibnamefont {Subasi}}, \bibinfo
  {author} {\bibfnamefont {L.}~\bibnamefont {Cincio}},\ and\ \bibinfo {author}
  {\bibfnamefont {P.~J.}\ \bibnamefont {Coles}},\ }\bibfield  {title} {\bibinfo
  {title} {Variational {Q}uantum {L}inear {S}olver},\ }\href
  {https://doi.org/10.22331/q-2023-11-22-1188} {\bibfield  {journal} {\bibinfo
  {journal} {{Quantum}}\ }\textbf {\bibinfo {volume} {7}},\ \bibinfo {pages}
  {1188} (\bibinfo {year} {2023})}\BibitemShut {NoStop}%
\bibitem [{\citenamefont {Turati}\ \emph {et~al.}(2024)\citenamefont {Turati},
  \citenamefont {Marruzzo}, \citenamefont {Dacrema},\ and\ \citenamefont
  {Cremonesi}}]{eni2024}%
  \BibitemOpen
  \bibfield  {author} {\bibinfo {author} {\bibfnamefont {G.}~\bibnamefont
  {Turati}}, \bibinfo {author} {\bibfnamefont {A.}~\bibnamefont {Marruzzo}},
  \bibinfo {author} {\bibfnamefont {M.~F.}\ \bibnamefont {Dacrema}},\ and\
  \bibinfo {author} {\bibfnamefont {P.}~\bibnamefont {Cremonesi}},\ }\href
  {https://arxiv.org/abs/2409.06339} {\bibinfo {title} {An empirical analysis
  on the effectiveness of the variational quantum linear solver}} (\bibinfo
  {year} {2024}),\ \Eprint {https://arxiv.org/abs/2409.06339} {arXiv:2409.06339
  [quant-ph]} \BibitemShut {NoStop}%
\bibitem [{\citenamefont {Bhatnagar}\ \emph {et~al.}(1954)\citenamefont
  {Bhatnagar}, \citenamefont {Gross},\ and\ \citenamefont
  {Krook}}]{Bhatnagar1954}%
  \BibitemOpen
  \bibfield  {author} {\bibinfo {author} {\bibfnamefont {P.~L.}\ \bibnamefont
  {Bhatnagar}}, \bibinfo {author} {\bibfnamefont {E.~P.}\ \bibnamefont
  {Gross}},\ and\ \bibinfo {author} {\bibfnamefont {M.}~\bibnamefont {Krook}},\
  }\bibfield  {title} {\bibinfo {title} {A model for collision processes in
  gases. i. small amplitude processes in charged and neutral one-component
  systems},\ }\href {https://doi.org/10.1103/PhysRev.94.511} {\bibfield
  {journal} {\bibinfo  {journal} {Phys. Rev.}\ }\textbf {\bibinfo {volume}
  {94}},\ \bibinfo {pages} {511} (\bibinfo {year} {1954})}\BibitemShut
  {NoStop}%
\bibitem [{\citenamefont {Sanavio}\ and\ \citenamefont
  {Succi}(2024)}]{Sanavio2024}%
  \BibitemOpen
  \bibfield  {author} {\bibinfo {author} {\bibfnamefont {C.}~\bibnamefont
  {Sanavio}}\ and\ \bibinfo {author} {\bibfnamefont {S.}~\bibnamefont
  {Succi}},\ }\bibfield  {title} {\bibinfo {title} {Lattice
  boltzmann–carleman quantum algorithm and circuit for fluid flows at
  moderate reynolds number},\ }\href {https://doi.org/10.1116/5.0195549}
  {\bibfield  {journal} {\bibinfo  {journal} {AVS Quantum Science}\ }\textbf
  {\bibinfo {volume} {6}},\ \bibinfo {pages} {023802} (\bibinfo {year}
  {2024})}\BibitemShut {NoStop}%
\bibitem [{\citenamefont {Tsemo}\ \emph {et~al.}(2024)\citenamefont {Tsemo},
  \citenamefont {Jayashankar}, \citenamefont {Sugisaki}, \citenamefont
  {Baskaran}, \citenamefont {Chakraborty},\ and\ \citenamefont
  {Prasannaa}}]{tsemo2024enhancingharrowhassidimlloydhhlalgorithm}%
  \BibitemOpen
  \bibfield  {author} {\bibinfo {author} {\bibfnamefont {P.~B.}\ \bibnamefont
  {Tsemo}}, \bibinfo {author} {\bibfnamefont {A.}~\bibnamefont {Jayashankar}},
  \bibinfo {author} {\bibfnamefont {K.}~\bibnamefont {Sugisaki}}, \bibinfo
  {author} {\bibfnamefont {N.}~\bibnamefont {Baskaran}}, \bibinfo {author}
  {\bibfnamefont {S.}~\bibnamefont {Chakraborty}},\ and\ \bibinfo {author}
  {\bibfnamefont {V.~S.}\ \bibnamefont {Prasannaa}},\ }\href
  {https://arxiv.org/abs/2407.21641} {\bibinfo {title} {Enhancing the
  harrow-hassidim-lloyd (hhl) algorithm in systems with large condition
  numbers}} (\bibinfo {year} {2024}),\ \Eprint
  {https://arxiv.org/abs/2407.21641} {arXiv:2407.21641 [physics.atom-ph]}
  \BibitemShut {NoStop}%
\bibitem [{\citenamefont {Berry}\ \emph {et~al.}(2006)\citenamefont {Berry},
  \citenamefont {Ahokas}, \citenamefont {Cleve},\ and\ \citenamefont
  {Sanders}}]{Berry2006}%
  \BibitemOpen
  \bibfield  {author} {\bibinfo {author} {\bibfnamefont {D.~W.}\ \bibnamefont
  {Berry}}, \bibinfo {author} {\bibfnamefont {G.}~\bibnamefont {Ahokas}},
  \bibinfo {author} {\bibfnamefont {R.}~\bibnamefont {Cleve}},\ and\ \bibinfo
  {author} {\bibfnamefont {B.~C.}\ \bibnamefont {Sanders}},\ }\bibfield
  {title} {\bibinfo {title} {Efficient quantum algorithms for simulating sparse
  hamiltonians},\ }\href {https://doi.org/10.1007/s00220-006-0150-x} {\bibfield
   {journal} {\bibinfo  {journal} {Communications in Mathematical Physics}\
  }\textbf {\bibinfo {volume} {270}},\ \bibinfo {pages} {359–371} (\bibinfo
  {year} {2006})}\BibitemShut {NoStop}%
\bibitem [{\citenamefont {Nielsen}\ and\ \citenamefont
  {Chuang}(2010)}]{Nielsen_Chuang_2010}%
  \BibitemOpen
  \bibfield  {author} {\bibinfo {author} {\bibfnamefont {M.~A.}\ \bibnamefont
  {Nielsen}}\ and\ \bibinfo {author} {\bibfnamefont {I.~L.}\ \bibnamefont
  {Chuang}},\ }\href {https://doi.org/https://doi.org/10.1017/CBO9780511976667}
  {\emph {\bibinfo {title} {Quantum Computation and Quantum Information: 10th
  Anniversary Edition}}}\ (\bibinfo  {publisher} {Cambridge University Press},\
  \bibinfo {year} {2010})\BibitemShut {NoStop}%
\bibitem [{\citenamefont {Childs}\ \emph {et~al.}(2017)\citenamefont {Childs},
  \citenamefont {Kothari},\ and\ \citenamefont {Somma}}]{HHL_improvement1}%
  \BibitemOpen
  \bibfield  {author} {\bibinfo {author} {\bibfnamefont {A.~M.}\ \bibnamefont
  {Childs}}, \bibinfo {author} {\bibfnamefont {R.}~\bibnamefont {Kothari}},\
  and\ \bibinfo {author} {\bibfnamefont {R.~D.}\ \bibnamefont {Somma}},\
  }\bibfield  {title} {\bibinfo {title} {Quantum algorithm for systems of
  linear equations with exponentially improved dependence on precision},\
  }\href {https://doi.org/10.1137/16M1087072} {\bibfield  {journal} {\bibinfo
  {journal} {SIAM Journal on Computing}\ }\textbf {\bibinfo {volume} {46}},\
  \bibinfo {pages} {1920} (\bibinfo {year} {2017})}\BibitemShut {NoStop}%
\bibitem [{\citenamefont {Wossnig}\ \emph {et~al.}(2018)\citenamefont
  {Wossnig}, \citenamefont {Zhao},\ and\ \citenamefont
  {Prakash}}]{PhysRevLett.120.050502}%
  \BibitemOpen
  \bibfield  {author} {\bibinfo {author} {\bibfnamefont {L.}~\bibnamefont
  {Wossnig}}, \bibinfo {author} {\bibfnamefont {Z.}~\bibnamefont {Zhao}},\ and\
  \bibinfo {author} {\bibfnamefont {A.}~\bibnamefont {Prakash}},\ }\bibfield
  {title} {\bibinfo {title} {Quantum linear system algorithm for dense
  matrices},\ }\href {https://doi.org/10.1103/PhysRevLett.120.050502}
  {\bibfield  {journal} {\bibinfo  {journal} {Phys. Rev. Lett.}\ }\textbf
  {\bibinfo {volume} {120}},\ \bibinfo {pages} {050502} (\bibinfo {year}
  {2018})}\BibitemShut {NoStop}%
\bibitem [{\citenamefont {Shende}\ \emph {et~al.}(2005)\citenamefont {Shende},
  \citenamefont {Bullock},\ and\ \citenamefont
  {Markov}}]{Shannon_decomposition}%
  \BibitemOpen
  \bibfield  {author} {\bibinfo {author} {\bibfnamefont {V.}~\bibnamefont
  {Shende}}, \bibinfo {author} {\bibfnamefont {S.}~\bibnamefont {Bullock}},\
  and\ \bibinfo {author} {\bibfnamefont {I.}~\bibnamefont {Markov}},\
  }\bibfield  {title} {\bibinfo {title} {Synthesis of quantum logic circuits},\
  }in\ \href {https://doi.org/10.1109/ASPDAC.2005.1466172} {\emph {\bibinfo
  {booktitle} {Proceedings of the ASP-DAC 2005. Asia and South Pacific Design
  Automation Conference, 2005.}}},\ Vol.~\bibinfo {volume} {1}\ (\bibinfo
  {year} {2005})\ pp.\ \bibinfo {pages} {272--275 Vol. 1}\BibitemShut {NoStop}%
\end{thebibliography}%

\appendix

\section{Carleman linearization formula \label{app:Carleman_formula}}

\subsection{Collision Operator Details}
For a second-order Carleman expansion, Eq.~\eqref{eq:collision_carleman} is obtained, where the collision expressions for $E_{ij}$ and $D_{ijk}$ are given by
\begin{equation}
    D_{ij} = (1 - \omega) \delta_{ij} + \omega w_i \left(1 + \frac{e_i \cdot e_j}{c_s^2}\right)
\end{equation}
and
\begin{equation}
   E_{ijk} = \frac{\omega w_i}{c_s^4} \left( e_i \cdot e_j \,e_i \cdot e_k - c_s^2 e_j \cdot e_k \right)\,,
\end{equation}
where $\omega$, $e_i$ and $w_i$ are described in the main text.

\subsection{Streaming Operator Details }
For non-boundary points, the streaming operator at first order is given by a matrix as follows:
\begin{equation}
\left\{ \begin{array}{cl}
S_{kl} = 1   & \text{ if } k = (n,i) \text{ and } l = (n - e_i,i) \\
S_{kl} = 0   & \text{otherwise}\\
\end{array} \right.  \,,
\end{equation}
where, in this notation, the index $k$ represents the row of the matrix associated with the lattice point $n$ and velocity $i$. The variable $l$ is analogous for the column.

The streaming formulation differs only at the boundaries (for details on different boundary conditions see App.~\ref{app:BC}):
\begin{itemize}
    \item PBCs: $S_{kl} = 1$ if $k = (n,i)$ and $l = (m,i)$, where $m$ is the periodic point of $n$ along the direction $i$.
    \item Bounce-back: for fluid component that scatters on the wall, $S_{kl} = 1$ if $k = (n, i)$ and $l = (n,\bar{i})$, where $\bar{i}$ indicates the index for $\bar{e_i} = -e_i$.
    \item Lid driven cavity: we generalize the formulation of the bounce-back. We have more contributions for the row $k=(n,\bar{i})$: $S_{kl} = 1 - 2 \frac{1}{c_s^2} v_{\text{lid}} \cdot e_i$ if $k = (n, \bar{i})$ and $l = (n,i)$; $S_{kl} = -2 \frac{1}{c_s^2} v_{\text{lid}} \cdot e_i$ with $l = (n - e_j, j)$ (if the $j$ components points the wall, we have to change in $l = (n, \bar{j})$). The second contribution give us that $\sum_i -2\frac{1}{c_s^2} v_{lid} \cdot e_i f_i=  -2\frac{1}{c_s^2} v_{lid} \cdot e_i \rho_w$.
\end{itemize}  

The evolution of the $g$ components is obtained by multiplying the $g$ with $S \otimes S$.

\section{CNOT gate upper bounds for the Hamiltonian simulation \label{app:numbergats}}

On real devices, the Hamiltonian simulation part of the HHL algorithm must be compiled using sets of native one- and two-qubit gates. The most popular native two-qubit gates are CNOT gates. For a generic $n$-qubit unitary, it is possible to estimate that, at maximum, we need an order of $4^n$ CNOT gates~\cite{Shannon_decomposition}. In our case, the Hamiltonian simulation requires implementing $n_c$ times a single controlled-real-time evolution $e^{itA}$ which requires a number of qubits of $n_i = 2 + \log_2(N_t + 1) + \log_2(Q L)$, where the factor 2 comes from two contributions: one qubits for the control one and the other for the Hermitization of $A$. Using the general upper bound, we can implement our Hamiltonian simulation with at most 
\begin{equation}
  G_{qc} =  n_c 4^{n_i} \sim 16 n_c L^2 Q^2 N_t^2 \text{ CNOT gates.}
\end{equation}

However, it can be shown that the number of CNOT gates grows linearly with $L$, at the cost of increasing the number of qubits. The new mapping describes the fluid distribution for each lattice point with a specific group of qubits. To show the linear dependence of the CNOT gate on the lattice points, we use the local structure of the Carleman LBM method. In other words, the final matrix in Eq.~\eqref{eq:Carleman_full} has a block structure where it repeats $L$ times the same small matrix $\textbf{a}$, which propagates the fluid components of nearby lattice points (8 neighboring points + itself)\footnote{At the boundary, the matrix $\textbf{a}$ is different, but this aspect can be ignored because the $\textbf{a}$ matrix at the boundary acts on $\leq Q$ lattice points as well}. 

Using the upper bound of the number of CNOT gates ($4^n$), we can compile the single small matrix $\textbf{a}$ with at most $\tilde{Q} = 4^{Q \lceil \log_2(Q) \rceil}$ CNOT gates. Then, we have to repeat the quantum circuit $L$ times to cover all lattice points. Hence, the number of gates $G_{qc}$ that implements the full Hamiltonian simulation becomes
\begin{equation}
\begin{split}
G_{qc} &\leq n_c\,L \tilde{Q} \, 4^{2 + \log_2(N_t + 1)} \\
       &\leq 16 n_c\,L \, \tilde{Q}  \tilde{N_t}^2,
\end{split}
\end{equation}
where $\tilde{N_t} = 2^{\lceil \log_2(N_t + 1) \rceil}$. However, in our estimation, $\tilde{Q}$ might be a large number $\sim 4^{36}$, but we expect the actual value of $\tilde{Q}$ to be very small because the Carleman matrix acts only onto $Q$ states. We expect that $\tilde{Q} \sim 4^{\log(Q)} \sim 4^4$ because this is the cost of doing the collision plus additional specific SWAP gates for the application of streaming. Hence, the compilation of the Hamiltonian simulation of the Carleman LBM method depends linearly on the lattice points and quadratically on the number of time steps.

Moreover, due to the specific encoding, we gain a further advantage in reinitializing the $b$ vector. Different lattice points are mapped to different qubits, in other words, the $b$ state is given by a tensor product of the lattice points. Therefore, the only cost is the reinitialization of fluid components for each lattice points with a cost of $\sim 4^Q\sim 256$ CNOT gates.

\section{Benchmark Systems and Associated Boundary Conditions\label{app:BC}}
We recall that the initial fluid distribution for the PBC and bounce-back is given by the Kolmogorov flow defined in Eq.~\ref{eq:kolmogorov}. Instead, the initial distribution for the lid-driven cavity is a steady fluid (computed from Eq.~\eqref{eq:kolmogorov} with $A_x=A_y=0$), except for the fluid next to the moving wall that has the same lid velocity $v_{lid}$. The actual walls are placed halfway between the boundary and the nearest lattice node.

\subsection{Squared Box with Periodic Boundary Conditions}
Here we consider a squared box of size $N_x \times N_y$ whose walls are modeled via periodic boundary conditions. 
Such conditions ensure that fluid exiting one side of the domain seamlessly re-enter from the opposite side, maintaining continuity across the boundaries. Operatively, PBCs are applied to the distribution functions, ensuring a consistent representation of the system. As example, for the left side of a domain, the distribution functions are updated as follows:
\begin{align*}
f_1(t_1,n_x, n_y) &= f_1(t,n_x + N_x, n_y), \\
f_5(t+1,n_x+1, n_y) &= f_5(t,n_x + N_x, n_y), \\
f_8(t+1,n_x-1, n_y) &= f_8(t,n_x + N_x, n_y).
\end{align*}

\subsection{Squared Box with Bounce-Back  Boundary Conditions \label{app:bounceback}}
Here we consider a squared box of size $N_x \times N_x$ whose walls are modeled via bounce-back boundary conditions. Such conditions imply that the fluid distributions elastically scatter against the wall. The final fluid distribution is described by the following equation:
\begin{equation}
    f_{\bar{i}}(t+1,n)=f_{i}(t,n) \,,
\end{equation}
where $\vec{\bar{i}}=-\vec{i}$.
As an example, the streaming of the left wall is given by
\begin{align*}
f_1(t,n_x, n_y) &= f_3(t,n_x, n_y), \\
f_5(t,n_x, n_y) &= f_7(t,n_x, n_y), \\
f_8(t,n_x, n_y) &= f_6(t,n_x, n_y).\\
\end{align*}

\subsection{Squared Lid-Driven Cavity\label{app:lid_driven}}
Here we consider a squared box of size $N_x \times N_x$ whose walls are modeled as in the bounce back case, except for the left wall that describes here a lid moving at a constant velocity $\vec{v}=(0,\,v_{lid})$ along the $y$ direction.

To implement the moving lid, the bounce-back conditions are generalized to Dirichlet conditions, where the fluid distribution hitting the lid evolves according to
\begin{equation}
    f_{\bar{i}}(t+1,n)=f_{i}(t,n)+ 2 \frac{\rho_w}{c_s^2} v_{lid} \cdot e_{i} \,.\label{eq:diriclet}
\end{equation}
This means adjusting the fluid distribution based on the velocity of the lid.
Conversely, the corner sites connecting the lid with the rest of the box walls are computed with the standard bounce-back conditions.

\section{Lattice size study}
\label{app:size_study}
\begin{figure*}[t]
\centering
\includegraphics[width=1.\textwidth]{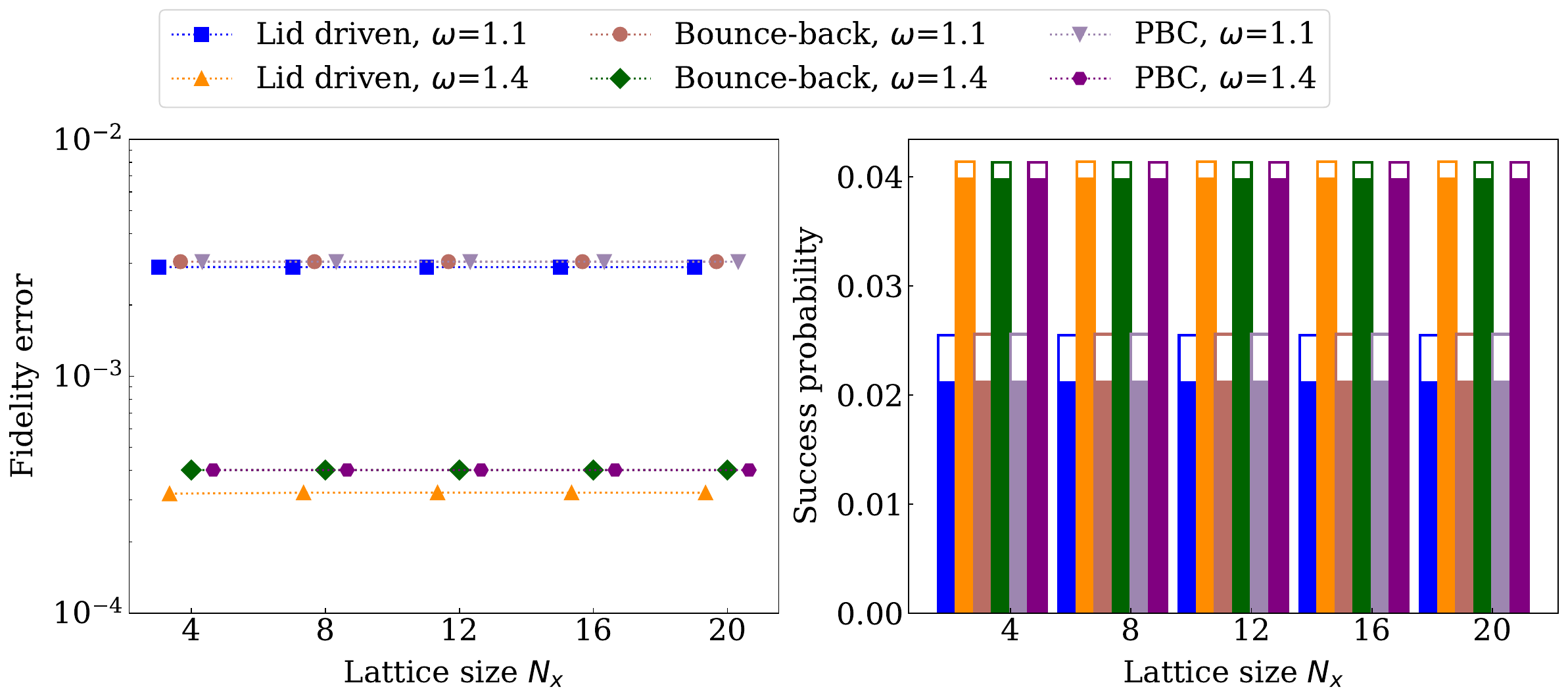}
\caption{
Left and right panels show the quantum fidelity and the success probability as a function of the lattice size $N_x$, respectively. In the left plot, the data points are slightly shifted for improved readability. In the right plot, filled symbols represent the success probability of measuring the clock in the
$\ket{0...0}$ state and ancilla in $\ket{1}$,  while empty symbols correspond to the probability of measuring the ancilla in $\ket{1}$ alone.  The bounce-back boundary conditions are evaluated for $t_0=0$, while the lid-driven case starts at $t_0=40$}
\label{fig:size_step1}
\end{figure*}

In this section we report the behavior of the various benchmark systems as a function of the lattice size, evolving for a single time step.
The number of clock qubits is fixed at $n_c=7$, while the relaxation frequency is set to $\omega=1.1,\,1.4$. In Fig.~\ref{fig:size_step1} we report the quantum fidelity and the success probability in the left and right panels, respectively. Empty bars correspond to measuring the ancilla in $\ket{1}$, while filled bars correspond to measuring the ancilla in $\ket{1}$ with the clock state in $\ket{0...0}$. 

\section{Success probability \label{app:success_vs_cp}}

\begin{figure}[t]
\centering
\includegraphics[width=1.\columnwidth]{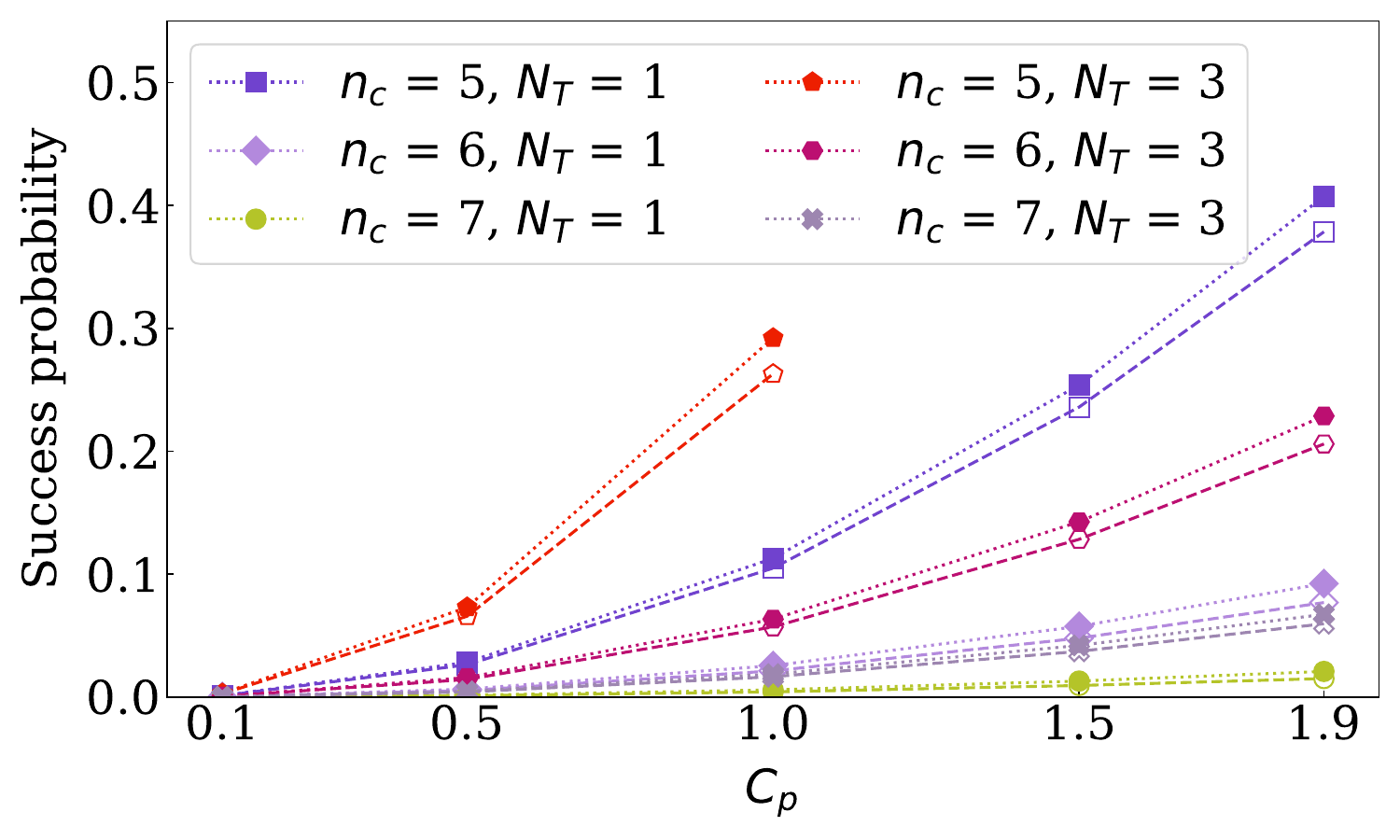}
\caption{Success probability of measuring the clock qubit in $\ket{0...0}$ state and the ancilla in $\ket{1}$ (colored symbols), and the probability of measuring the ancilla in the $\ket{1}$ state (empty symbols) as a function of angle parameters $C_p$ (see Eq.~\ref{eq:Cry_angles}). Simulations are carried out using a $8\times 8$ lattice with bounce-back conditions, various numbers of time steps $N_t$ for the time evolution, and various numbers of clock qubits $n_c$.
}
\label{fig:success_probability_vs_C_prob}
\end{figure}

We also study how the success probability depends on the coefficient $C_p$, which multiplies the angles in the controlled-$R_y$ gates. Figure~\ref{fig:success_probability_vs_C_prob} displays the success probability as a function of $C_p$ for a $8\times8$ lattice, with a bounce-back boundary condition, $\omega=1.1$, $N_t=1,3$, and $n_c=5,6,7$. The results show a quadratic growth along with $C_p$. 
Additionally, the success probability decreases with increasing $n_c$ but increases with evolution over multiple time steps. 

Note that we could not implement the HHL algorithm for the case $[n_c=6,N_t=3]$ for $C_p>1$ because the smallest binary eigenvalue of $A$ is $1$, so we cannot compute the controlled-$R_y$ angles because $\theta=\arcsin\left(\frac{C_p}{1}\right)$.

\end{document}